\documentclass[12pt, draftclsnofoot, onecolumn]{IEEEtran}
\usepackage{amssymb}
\usepackage{amsmath}
\usepackage{amsfonts}
\usepackage{eucal}
\usepackage{bbm}
\usepackage{amsfonts}
\usepackage[dvips]{graphicx}
\usepackage{graphicx}
\usepackage{amsmath}
\usepackage{fancyhdr}
\usepackage{stfloats}
\usepackage{multirow}
\usepackage{algorithm}
\usepackage{algorithmic}
\usepackage{cite}
\usepackage[bookmarks=false]{}
\usepackage{amsthm}
\usepackage{algorithm}
\usepackage{algorithmic}
\usepackage{mathbbol}
\usepackage{amsfonts}
\usepackage{graphicx}
\usepackage{amsmath}
\usepackage{amssymb}
\usepackage{latexsym}
\usepackage{graphicx}
\usepackage{cite}
\usepackage{url}
\usepackage{stfloats}
\usepackage{mathrsfs}
\usepackage{setspace}
\usepackage{booktabs}
\usepackage {latexsym}
\usepackage {graphicx}
\usepackage {cite}
\usepackage {color}
\usepackage {url}
\usepackage {stfloats}
\usepackage {mathrsfs}
\theoremstyle{plain}

\usepackage{cite}
\usepackage{setspace}
\ifCLASSINFOpdf
\usepackage[pdftex]{graphicx}
\else
\fi
\usepackage[tight,footnotesize]{subfigure}
\usepackage{colortbl}
\usepackage{float}
\usepackage{lipsum}

\makeatletter
\newenvironment{breakablealgorithm}
  {% \begin{breakablealgorithm}
   \begin{center}
     \refstepcounter{algorithm}% New algorithm
     \hrule height.8pt depth0pt \kern2pt% \@fs@pre for \@fs@ruled
     \renewcommand{\caption}[2][\relax]{% Make a new \caption
       {\raggedright\textbf{\ALG@name~\thealgorithm} ##2\par}%
       \ifx\relax##1\relax % #1 is \relax
         \addcontentsline{loa}{algorithm}{\protect\numberline{\thealgorithm}##2}%
       \else % #1 is not \relax
         \addcontentsline{loa}{algorithm}{\protect\numberline{\thealgorithm}##1}%
       \fi
       \kern2pt\hrule\kern2pt
     }
  }{% \end{breakablealgorithm}
     \kern2pt\hrule\relax% \@fs@post for \@fs@ruled
   \end{center}
  }
\makeatother

\begin{document}
%\pagenumbering{gobble}% Remove page numbers (and reset to 1)
\clearpage
%\maketitle
%Title.
% ------
\title{\huge A Generalized Dimming Control Scheme for Visible Light Communications}
\author{\IEEEauthorblockN{Yang Yang, \emph{Member, IEEE}, Congcong Wang,
Chunyan Feng, \emph{Senior Member, IEEE}, Caili Guo, \emph{Senior Member, IEEE}, Julian Cheng, \emph{Senior Member, IEEE} and Zhimin Zeng %\\\IEEEauthorrefmark{1}%Beijing Key Laboratory of Network System Architecture and Convergence, School of
%%Information and Communication Engineering, Beijing University of Posts and Telecommunications, Beijing 100876, China
%\\\IEEEauthorrefmark{2}%Virginia Polytechnic Institute and State University, Blacksburg, USA
%\\
} \thanks{Y. Yang, C. Wang C. Feng and Z. Zeng are with the Beijing Key Laboratory of Network System Architecture and Convergence, School of Information and Communication Engineering, Beijing University of Posts and Telecommunications, Beijing 100876, China (e-mail: yangyang01@bupt.edu.cn; wcongcong@bupt.edu.cn; cyfeng@bupt.edu.cn; zengzm@bupt.edu.cn).}
\thanks{C. Guo is with Beijing Laboratory of Advanced Information Networks, School of Information and Communication Engineering, Beijing University of Posts and Telecommunications, Beijing 100876, China (e-mail: guocaili@bupt.edu.cn).}
\thanks{J. Cheng is with the Faculty of Applied Science, School of Engineering, The University of British Columbia, Kelowna, BC V1V 1V7, Canada (e-mail: julian.cheng@ubc.ca).}}
\maketitle
\thispagestyle{empty}
\vspace{0cm}
\begin{abstract}
A novel dimming control scheme, termed as generalized dimming control (GDC), is proposed for visible light communication (VLC) systems.
The proposed GDC scheme achieves dimming control by simultaneously adjusting the intensity of transmitted symbols and the number of active elements in a space-time matrix.
Both the indices of the active elements in each space-time matrix and the modulated constellation symbols are used to carry information.
Since illumination is deemed as the prior task of VLC, an incremental algorithm for index mapping is proposed for achieving target optical power and uniform illumination.
Next, GDC having the optimal activation pattern is investigated to further improve the bit-error rate (BER) performance.
In particular, the BER performance of GDC is analyzed using the union bound technique.
Based on the analytical BER bound, the optimal activation pattern of GDC scheme with the minimum BER criterion (GDC-MBER) is obtained by exhaustively searching all conditional pairwise error probabilities.
However, since GDC-MBER requires high search complexity, two low-complexity GDC schemes having the maximum free distance criterion (GDC-MFD) are proposed. The first GDC-MFD scheme, coined as GDC-MFD1, reduces the computational complexity by deriving a lower bound of the free distance based on Rayleigh-Ritz theorem. Based on the time-invariance characteristics of the VLC channel, GDC-MFD2 is proposed to further reduce the required computation efforts.
Simulation and numerical results show that GDC-MBER, GDC-MFD1 and GDC-MFD2 have similar BER performance, and they can achieve 2 dB performance gains over conventional hybrid dimming control scheme and 7 dB performance gains over digital dimming control schemes.
\end{abstract}

\vspace{-0.cm}
\section{Introduction}
\label{sec:intro}
\IEEEPARstart{L}{ight} emitting diodes (LEDs) have attracted great attention in recent years due to their long life expectancy, high tolerance to humidity and low power consumption \cite{indoorVLC}. Based on the popularity of LEDs, an optical wireless access technology known as visible light communication (VLC) has gained increasing attentions. Compared with the conventional radio-frequency based technologies, VLC has low impact on human health, non-electromagnetic interference \cite{eDCOOFDM,YangFair}, and can offer high data-rate communication services \cite{Cachinginthesky,WangSecurity}.
Since VLC involves both communication and illumination, it is significant to investigate dimming control schemes, which aims at improving communication performance under the constraints of illumination.

\subsection{Related Works}
There is a plethora of prior art on dimming control schemes for VLC \cite{theStateOfResearch,YangAnEnhanced,amPAM,LEDcolorShift,ElgalaReverse,PWMdimming1,OOKDimming,DDr1,FECcodimg,modifiedreedmullercodes,spatialmodulation1,spatialmodulation2,SptialDimming}. Generally, the dimming control schemes can be classified into three types: analogue dimming (AD) \cite{theStateOfResearch,YangAnEnhanced,amPAM,LEDcolorShift}, digital dimming (DD) \cite{ElgalaReverse,PWMdimming1,OOKDimming,DDr1,FECcodimg,modifiedreedmullercodes}, and spatial dimming (SD) \cite{spatialmodulation1,spatialmodulation2,SptialDimming}. The AD technique is the simplest approach in dimming control, which controls optical power by adjusting the amplitude of the current \cite{theStateOfResearch}. It is relatively easy to implement and the luminous intensity is reduced proportionally to the current.
For instance, a direct current (DC) offset was added \cite{YangAnEnhanced} to adjust the amplitude of the current. In \cite{amPAM}, the authors proposed to adjust the dimming level by changing the intensity levels of the pulse amplitude modulation (PAM) symbols.
%Although AD is simple and cost-effective, when the dimming level is high or low, the electrical power of the signals is prone to be small to avoid the clipping noise caused by the dynamic range of LEDs. Hence, the BER performance of AD is degraded at high or low dimming levels due to the limited available electrical power.

%
In contrast, DD technique achieves dimming control by adjusting the duty cycle of the transmitted signals. For instance, pulse width modulation (PWM) was used \cite{PWMdimming1}, and the pulse width was varied proportionally to acquire the desired dimming level.
The dimming levels can also be adjusted by controlling the ratio of non-zero codes of the transmitted signals \cite{OOKDimming}. Since the DD technique reduces light intensity more linearly than the AD technique, it reduces the risk of chromaticity shifts.

%Although the DD scheme can achieve wide range dimming control, its spectral efficiency is limited by the duty cycle. As discussed in \cite{DDr1}, when a fixed modulation order is used, the spectral efficiency of DD decreases when the required dimming level is at high or low level. Therefore, to maintain the same data rate at various dimming levels, the modulation order must increase as the dimming level becomes high or low.

Furthermore, SD technology was proposed to achieve dimming control by adjusting the number of glared LEDs  \cite{spatialmodulation1,spatialmodulation2}, and thus a target dimming level can be achieved without altering the signal form. Specifically, the number of glared LEDs increases with the dimming level. Since more glared LEDs imply more diversity gains that a SD system can achieve, the BER performance of SD improves as the dimming level increases \cite{SptialDimming}.

%To improve the data transmission rate, SD was proposed in \cite{SptialDimming}. In SD, the number of glared LEDs was adjusted to achieve dimming control, and thus a target dimming level can be achieved without altering the signal form. However, the number of glared LEDs decreases as the decrease of dimming level. Since more glared LEDs imply more diversity gains that a SD system can achieve, the BER performance of SD improves as the dimming level increases \cite{SptialDimming}.

%In general, we can conclude that each type of dimming control scheme has certain drawbacks.
From the above analysis, we can observe that the electrical power constraint of AD, the duty cycle constraint of DD and the diversity gain variation of SD can significantly affect the communication performance under the illumination constraints. That implies the balance between the dimming control and the communication performance in VLC requires further study.
In particular, it is interesting to investigate the optimal signal form that can minimize the communication performance loss. Or, equivalently, it is of interest to develop a generalized dimming control scheme that can efficiently incorporate the existing dimming control schemes and dynamically optimize the signal form to achieve the optimal trade-off between the communication and illumination aspects of VLC.
Recently, several hybrid dimming control schemes have been proposed \cite{HybridDimming,HDr1,HDr2}. However, they have certain limitations.
For instance, SD and AD methods were combined \cite{HybridDimming}, in which dimming control is achieved by simultaneously adjusting the number of active LEDs and the amplitude of the current.
However, the time resource (i.e. the duty cycle of the signal) is under-explored for dimming control due to the limitation of the signal form  \cite{HybridDimming}.

\subsection{Contributions}
The main contribution of this paper is a generalized dimming control (GDC) scheme that enables the joint use of the amplitude of the signals, duty cycle of the signals and the number of activated LEDs for dimming control. To the authors' best knowledge, this is the first generalized dimming control scheme that exploits all the three available resources in VLC for dimming control. The key contributions include the followings:
\begin{itemize}
\item We propose a generalized signal form for dimming control in VLC systems, in which both the intensity of the transmitted symbols and the number of active elements in a space-time matrix can be adjusted for dimming control. In particular, the duty cycle of the signals transmitted by an LED can be controlled by the active elements in a row of the designed space-time matrix; the number of active LEDs at a time slot can be controlled by the active elements in a column of the space-time matrix; and the intensity of each elements in the space-time matrix is controlled by a DC bias. In this way, all the AD, DD and SD schemes are incorporated in the proposed GDC scheme.
\item We first study the desired GDC signal form to satisfy certain optical power and uniform illumination constraints. In particular, an efficient incremental algorithm for index mapping is proposed to map bits into appropriate activation patterns\footnote{In this work, we define an activation pattern of a space-time matrix as a given realization of the matrix with $N_S$ non-zero elements. Note that the non-zero elements in the space-time matrix imply that the corresponding LED is activated at the time slot.} with uniform illumination. With a given activation pattern, the optical power of GDC is then investigated, and the optimal signal form of each matrix element for dimming control is studied.
\item Since multiple activation patterns can satisfy the illumination constraint, GDC having the optimal set of activation patterns is studied for further BER performance enhancement. In particular, the BER performance of GDC is analyzed and GDC having the minimum BER certerion (GDC-MBER) is first obtained by calculating the theoretical upper bound of BER. Since GDC-MBER requires exhaustively searching all conditional pairwise error probabilities (CPEPs) to calculate BER, two efficient suboptimal GDC designs with the maximum free distance criterion, i.e. GDC-MFD1 and GDC-MFD2, are then proposed. GDC-MFD1 reduces the computational complexity by deriving a lower bound of the free distance based on Rayleigh-Ritz theorem. GDC-MFD2 further reduces the computation complexity by utilizing the time-invariance characteristic of the VLC channel.
\end{itemize}
Simulation and analytical results show that the BER performance of  GDC-MFD1 and GDC-MFD2  approaches that of GDC-MBER at most dimming levels. In addition, all the three GDC schemes can achieve 2 dB and 8 dB performance gains over conventional hybrid dimming control scheme and digital dimming control scheme, respectively, when the dimming level is 50$\%$ and the BER is at $10^{-3}$.

The remainder of this paper is organized as follows. In Section II, we introduce the system model of this work. Section III presents the details of GDC. In Section IV, both the minimum BER and the maximum free distance criterions are proposed to optimize the activation pattern of GDC for further improving the communication reliability. Simulation and numerical results are presented and discussed in Section V. Finally, Section VI draws some important conclusions.

\emph{Notations:} ${\rm{E}}(\cdot)$, $\left\lfloor  \cdot  \right\rfloor$ and $\left\lceil  \cdot  \right\rceil$ denote the expectation operator, the floor and ceiling operators, respectively. The notation ${\left\|  \cdot  \right\|_F}$ represents the Frobenious norm of a vector or a matrix; ${\left(  \cdot  \right)^{\rm{T}}}$ is the transpose of a matrix/vector; ${\rm{C}}(\cdot ,\cdot )$ is the combination function; ${\rm{P}}(\cdot)$ is the probability function; ${\rm{Card}}( \cdot )$ represents the cardinality of a given set.

\vspace{-0.cm}
\section{System Model}
\label{sec:system}
\begin{figure}[H]
\setlength{\abovecaptionskip}{-0.2cm}
\centering
\includegraphics[width=0.7\textwidth]{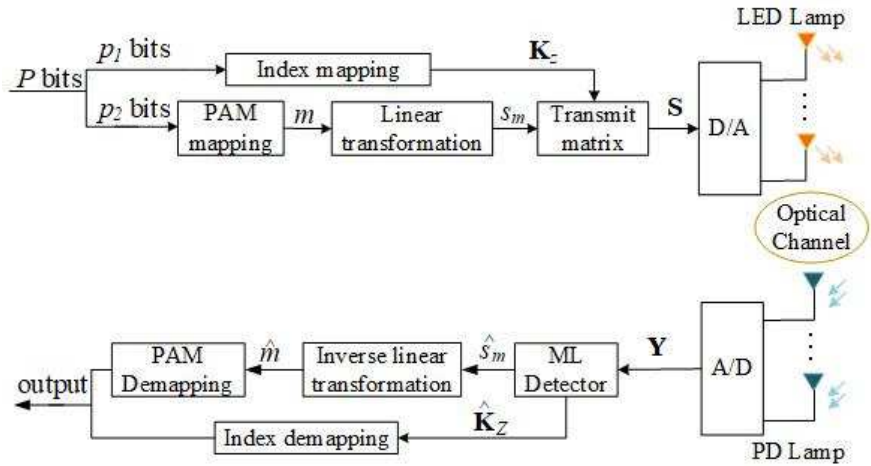}
\caption{Block diagram of the multi-LED VLC system with index modulation in spatial and time domains.}
\label{fig:system}
\vspace{-0.6cm}
\end{figure}

In this section, both the block diagram of the multi-LED VLC system and the optical wireless channel model are described. As shown in Fig. \ref{fig:system}, a multi-LED VLC system is employed, where the transmitter is equipped with $N_{t}$ LEDs and the receiver is equipped with $N_{R}$ photo-detectors (PDs). In this work, an intensity modulation direct detection system is considered and $M$-ary pulse amplitude modulation ($M$-PAM) is employed.

At the transmitter, the information is transmitted in terms of the space-time matrix, each carrying $P$ bits. The $P$ bits can further be divided into two parts.
The first part contains $p_1$ bits, which are mapped to an index matrix to select the active elements in the space-time matrix. The detailed index mapping criterion will be described in Section III. Besides, the second part contains ${p_2} = P - {p_1}$ bits, which are modulated by a certain modulation scheme. Note that the proposed GDC scheme does not impose specific constraint on the use of modulation scheme, and in this work, we employ $M$-PAM having amplitude levels $ m \in \{ 1,2, \cdots ,M\}$ \cite{LEDcolorShift}.

In VLC, all the transmitted symbols should be restricted to the dynamic range of the LEDs. Otherwise, the signals will be clipped, resulting in clipping noise. To this end, a linear transformation is conducted to fit the $M$-PAM signals for transmission
\begin{eqnarray}\label{signalform}
{s_m}=\lambda m + {B_{\rm{L}}},\;\;{I_{\rm{L}}} < {s_m} \le {I_{\rm{H}}}
\end{eqnarray}
where $\lambda$ is the scaling factor, $B_L$ is the DC bias, ${I_{\rm{L}}}$ and ${I_{\rm{H}}}$ are the minimum and maximum currents allowed for the LED, respectively.

In VLC systems, the channel gain can be divided into the line of sight (LOS) part and the non-LOS part \cite{indoorVLC}. We assume that the optical wireless channel only has the LOS part in this work, which contains most parts of the transmitted energy. Then, the channel gain between the $n$-th ($1 \le n \le {N_t}$) LED and the $r$-th ($1 \le r \le {N_R}$) PD can be calculated as
\begin{eqnarray}
\label{cch}
{h_{r,n}} = \left\{ {\begin{array}{*{20}{l}}
{\frac{{(l + 1)A}}{{2\pi {d^2}}}g(\psi ){{\cos }^l}(\phi )\cos (\psi )},&0 < \psi  \le {\Psi _{\rm{C}}}\\
{0},&{\psi  > {\Psi _{\rm{C}}}}
\end{array}} \right.
\end{eqnarray}
where $l =  - \frac{{\ln 2}}{{\ln (\cos {\Phi _{1/2}})}}$ is the order of the Lambertian emission, ${\Phi _{1/2}}$ is the semi-angle of LEDs at the half illumination power value, $A$ is the detector area, $d$ is the distance between the $n$-th LED and the $r$-th PD, and $g(\psi )$ denotes the gain of optical concentrator
\begin{eqnarray}
g(\psi ) = \left\{ \begin{array}{l}
\frac{{{n_r}^2}}{{{{\sin }^2}{\Psi _{\rm{C}}}}},\hspace{5pt}0 < \psi  \le  {\Psi _{\rm{C}}}\\
0,\hspace{28pt}\psi  > {\Psi _{\rm{C}}}
\end{array} \right.
\end{eqnarray}
where ${{n_r}}$ is the refractive index, ${\Psi _{\rm{C}}}$ is the receiver field of vision semi-angle.
In (\ref{cch}), $\phi $ and $\psi$ are the angle of emergence with respect to the transmitter axis and the angle of incidence with respect to the receiver axis, respectively.

At the receiver, optical signals are first converted into electronic signals by PDs. Then, the received signal vector at the $j$-th ($1 \le j \le T$) time slot ${{\bf{Y}}_j} \in {{\cal{R}}^{{N_R} \times 1}}$ can be obtained as
\begin{eqnarray}\label{simmodel}
{\bf{Y}}_j = {{\bf{HK}}_z^j}{s_m} + {\bf{W}}_j = {\bf{HS}}_j + {\bf{W}}_j
\end{eqnarray}
where ${\bf{H}} \in {{\cal{R}}^{{N_R} \times {N_t}}}$ is the channel matrix,  ${\bf{K}}_z^j \in {{\cal{R}}^{{N_t} \times 1}}$ is the $j$-th column of the index matrix with the $z$-th activation pattern ${\bf K}_z \in {{\cal{R}}^{{N_t} \times T}}$, ${\bf{W}}_j \in {{{\cal{R}}^{{N_R} \times 1}}}$ is the noise vector, and ${\bf S}_j = {\bf{K}}_z^j{s_m}$ is the $j$-th column of the transmitted matrix ${\bf S} \in {{\cal{R}}^{{N_t} \times T}}$. In addition, the elements of ${\bf{W}}_j$ follow the independent and identically (i.i.d.) Gaussian distribution with zero-mean and variance ${N_0}$.
As the positions of the transmitters and the receivers are fixed during the transmission, the channel gain matrix during each time slot is equal to ${\bf{H}}$. Therefore, the system model can be expressed as
\begin{eqnarray}
{\bf{Y}} = {\bf{H}}{\bf{K}}_z{s_m} + {\bf{W}} = {\bf{H}}{\bf{S}} + {\bf{W}}
\end{eqnarray}
where ${\bf{Y}} \in {{{\cal{R}}^{{N_R} \times T}}}$ consists of all the symbols that ${N_R}$ PDs receive in $T$ time slots, ${\bf{W}} \in {{{\cal{R}}^{{N_R} \times T}}}$ is the noise matrix, and ${\bf{S}}$ is generated by combining an index matrix ${{\bf{K}}_z}$ with $s_m$ (i.e. ${\bf{S}}  = {{\bf{K}}_z}{s_m}$), which consists of all the transmit symbols that ${N_t}$ LEDs transmit in $T$ time slots.
Furthermore, a maximum likelihood (ML) detector is used to detect the received signals for GDC. The ML detection criterion is formulated as
\begin{eqnarray}
\begin{split}
{\left( {{{\widehat {\bf{K}}}_z},{{\widehat s}_m}} \right)_{{\rm{ML}}}} = \mathop {\arg \min }\limits_{{{\bf{K}}_z} \in {\mathbb K},1 \le m \le M} {\left\| {{\bf{Y}} - {\bf{HS}}} \right\|_F^2}
\end{split}
\end{eqnarray}
where ${\mathbb K}$ denotes the set of index matrices with all possible activation patterns, ${\widehat {\bf{K}}_z}$ and ${{\widehat s}_m}$ are the estimates of ${\bf{K}}_z$ and $s_m$, respectively. Based on the detected ${\widehat {\bf{K}}_z}$, we can easily obtain the decimal number $z$ by index demapping. Applying $z$ into a decimal-to-binary converter, we can recover the $p_1$ bits. Furthermore, based on the detected ${{\widehat s}_m}$, we can obtain the corresponding PAM symbol according to the inverse linear transformation of (\ref{signalform}).

\section{Proposed GDC Scheme}
\label{sec:STIM Scheme}
This section elaborates the design of GDC, which mainly consists of the space-time matrix design and the transmitted signal design. First, we review a combinatorial based index mapping algorithm for the space-time matrix design, which can modulate bits into different space-time matrices in GDC \cite{YesilkayaOptical}.
However, since different activation patterns of GDC can result in uneven illumination, based on the combinatorial index mapping algorithm, a low-complexity incremental index mapping algorithm that achieves uniform illumination is then proposed.
Finally, we finish the signal design of GDC. In particular, the optimal PAM signal form that minimizes BER is investigated under the optical power and data transmission rate constrains.

\vspace{-0.2cm}
\subsection{Combinatorial Based Index Mapping}
In this subsection, we review a combinatorial based index mapping algorithm that transforms bits into an activation pattern of the space-time matrix in GDC \cite{combinemap}. Assume $N_S$ out of ${N_t}T$ elements in a space-time matrix are activated for transmission, while the rest $\left( {{N_t}T - {N_S}} \right)$ elements are set to zero.
Therefore, there are ${{\rm{C}}({N_t}T,{N_S})}$ possibilities for the activation patterns of the space-time matrix for a given $N_S$.
Note that ${{\rm{C}}({N_t}T,{N_S})}$ denotes the number of combinations of selecting $N_S$ out of $N_t T$ elements. That means ${p_1} = \left\lfloor {{{\log }_2}\left( {{\rm{C}}({N_t}T,{N_S})} \right)} \right\rfloor$ bits can be carried through the variation of the activation patterns of the space-time matrices. Thus, the number of bits carried by each space-time matrix can be denoted as
\begin{eqnarray}\label{SE}
P = {p_1} + {p_2} = \left\lfloor {{{\log }_2}\left( {{\rm{C}}({N_t}T,{N_S})} \right)} \right\rfloor + {\log _2}M.
\end{eqnarray}

Both the look-up table method and the combinatorial method can be used to map each $p_1$ bits to a unique activation pattern \cite{combinemap}.
For given ${N_S},{N_t}$ and $T$, a look-up table of size $\left( {{2^{\left\lfloor {{{\log }_2}\left( {{\rm{C}}({N_t}T,{N_S})} \right)} \right\rfloor }}} \right)$ is required. Obviously, the size of look-up table increases exponentially with ${\left\lfloor {{{\log }_2}\left( {{\rm{C}}({N_t}T,{N_S})} \right)} \right\rfloor }$.
Since  GDC may require large values of ${N_S}$ and ${N_t}T$ for precise dimming accuracy and wide dimming range, manually designing a look-up table is not desirable for GDC.

In contrast, the combinatorial method can be implemented in a simpler way. The combinatorial method provides a one-to-one mapping rule between a natural number ${\rm{0}} \le z \le {\rm{C}}\left( {{N_t}T,{N_S}} \right) - 1$ and a descending sequence $\left\{ {{c_{{N_S}}},{c_{{N_S} - 1}}, \cdots ,{c_1}} \right\}$ as follows \cite{combinemap}
\begin{equation}\label{map}
z = {\rm{C}}({c_{{N_S}}},{N_S})+{\rm{C}}({c_{{N_S}-1}},{N_S-1})+ \ldots + {\rm{C}}({c_{1}},{1})
\end{equation}
where ${c_{{N_S}}} > {c_{{N_{S}- 1}}} >  \ldots  > {c_1}$.
Furthermore, the descending sequence  $\left\{ {{c_{{N_S}}},{c_{{N_S} - 1}}, \cdots ,{c_1}} \right\}$ are used to denote the active elements in the activation patterns.
Let ${{\bf{K}}_z}$ be the matrix denoting the $z$-th activation pattern, in which $k_z^{n,j} \in \left\{ {0,1} \right\}$ is the element at the $n$-th row and the $j$-th column of ${{\bf{K}}_z}$. Note that $k_z^{n,j} = 1$ denotes that the $n$-th LED is ``active'' during the $j$-th time slot, while $k_z^{n,j} = 0$ denotes that the $n$-th LED is ``silent'' during the $j$-th time slot.
Furthermore, the mapping rule between the descending sequence $\left\{ {{c_{{N_S}}},{c_{{N_S} - 1}}, \cdots ,{c_1}} \right\}$ and the index matrix ${\bf{K}}_z$ is as follows: for $a = 1, \cdots ,{N_S}$, if and only if $n = \left\lfloor {\frac{{{c_a}{\rm{ + 1}}}}{T}} \right\rfloor$ and $j = \left\lfloor {\frac{{{c_a}{\rm{ + 1}}}}{N_t}} \right\rfloor$, we have $k_z^{n,j} = 1$, while the remaining elements in matrix ${{\bf{K}}_z}$ are set to zero. Since $2^{p_1}$ combinations are sufficient to carry ${p_1}$ bits, $\left( {{\rm{C}}({N_t}T,{N_S}) - {2^{{p_1}}}} \right)$ extra activation patterns need to be discarded. Note that the activation probabilities of LEDs vary with the activation patterns, and different activation probabilities of each LED can result in uneven illumination distribution. To achieve uniform illumination, we consider an incremental algorithm for index mapping in the next subsection.

\subsection{Index Mapping Algorithm with Uniform Illumination}
In this work, we assume that the locations of LEDs are predetermined, and the uniform illumination requirement can be satisfied once all the LEDs have the same (or almost the same) activation probabilities. Based on this preliminary, we can select the optimal set of activation patterns for a given $N_S$.
%The transmitted symbol vector can be expressed as ${\bf{S}} = {{\bf{K}}_z}{s_m}$ for $z \in \left\{ {0,1, \cdots ,{\rm{C}}\left( {{N_t}T,{N_S}} \right) - 1} \right\}$ and $m \in \left\{ {1,2, \cdots ,M} \right\}$. Since only ${2^{{p_1}}} \le {\rm{C}}({N_t}T,{N_S})$ combinations are used as legal active patterns, and thus several extra active patterns need to be discarded.
%For example, when ${N_t}T = 8,{N_S} = 4$, we can obtain ${p_1} = {\log _2}\left\lfloor {{\rm{C}}({N_t}T,{N_S})} \right\rfloor  = 6$, ${{\rm{2}}^{{p_1}}} = 64$ and ${\rm{C}}({N_t}T,{N_S}) = {\rm{70}}$. Therefore, six extra activate patterns need to be discarded.
%
Intuitively, the desired active pattern can be obtained by an exhaustive search. In particular, the optimal set of activation patterns can be obtained by finding the ${2^{{p_1}}}$ activation patterns that minimize the variance of the activation probability of all LEDs. The problem can be formulated as
\begin{align}\label{ilu_problem}
\mathbb{O} = {\left\{ {{{{\bf{K}} }_{{o_1}}},{{{\bf{K}} }_{{o_2}}}, \ldots ,{{{\bf{K}} }_{{o_{{2^{p1}}}}}}} \right\}}
 &= {\mathop {\arg \min }\limits_{\left\{ {{{\bf{K}}_{{o_1}}},{{\bf{K}}_{{o_2}}}, \ldots ,{{\bf{K}}_{{o_{{2^{p1}}}}}}} \right\}} {\sigma ^2}\left\{ {\sum\limits_{q = 1}^{{2^{p1}}} {u_{{o_q}}^1} ,\sum\limits_{q = 1}^{{2^{p1}}} {u_{{o_q}}^2} , \cdots ,\sum\limits_{q = 1}^{{2^{p1}}} {u_{{o_q}}^{{N_t}}} } \right\}}\\
&{\rm{s.\;t.}}\;\;{\rm{Card}}\left( {\mathbb{O}} \right) = 2^{p_1},\tag{\theequation a}\\
&\;\;\;\;\;\;\;\;\;\mathbb{O} \subset  \mathbb{K} \tag{\theequation b}
\end{align}
where for the $o_q$-th activation pattern ${\bf{K}}_{o_q}$, $u_{{o_q}}^n = \sum\limits_{j = 1}^T {k_{{o_q}}^{n,j}} ,n = 1,2, \cdots {N_t}$ counts the cumulative number of active time slots for the $n$-th LED. In addition, ${\sigma ^2}\left\{ \bf {a} \right\}$ is the variance of the elements in vector $\bf a$, $\mathbb{O}$ is the desired set of activation patterns that achieves uniform illumination, and $\mathbb{K}$ is the set of all possible activation patterns.
The search complexity of conventionally ES is ${\cal{O}} \left({\rm{C}}\left( { {\rm{Card}}\left( {\mathbb{K}} \right) , {\rm{Card}}\left( {\mathbb{O}} \right) } \right)\right)$, which can be extremely complicated when the value of ${\rm{Card}}\left( {\mathbb{K}} \right)$ is large.

To circumvent this challenge, we propose a low-complexity incremental algorithm for index mapping.
The basic idea of the incremental algorithm can be described as follows.
First, the combinatorial mapping method is used to obtain all possible activation patterns of the space-time matrix, and then we use ${{\bf{U}}_z} \in {{\cal{R}}^{{N_t} \times 1}}$ to record the cumulative activation times of each LED of the $z$-th possible activation pattern. Second, we divide all the activation patterns into $N_t$ subsets. The subset is divided as follows: for the $n$-th subset, the cumulation activation numbers of the first $(n-1)$ LEDs are $0$.
Third, we sequentially select the activation patterns in the $n$-th subset into a set $\mathbb{O}$ until the activation number of the $n$-th LED equals to $\left\lceil {\frac{{{2^{{p_1}}}{N_S}}}{{{N_t}}}} \right\rceil$. Finally, a compensation process is designed for the case ${\rm Card} \left( \mathbb{O} \right)  \ne 2^{p_1}$ to minimize the variance of the activation probability of all LEDs. The detail of our proposed incremental algorithm is shown in \textbf{Algorithm} \ref{algorithm1}.

\vspace{0.5cm}
\begin{breakablealgorithm}
\caption{Incremental algorithm for index mapping.}
\small
\begin{algorithmic}[1]
\begin{spacing}{1.0}
\STATE \textbf{Input:} $N_t,\;T,\;N_S$.
\FOR {$z$ = 0 $ \to $ ${\rm{C}}({N_t}T,{N_S})-1$}
\STATE Map the decimal integer $z$ to the descending sequence $\left\{ {{c_{{N_S}}},{c_{{N_S} - 1}}, \cdots ,{c_1}} \right\}$ according to (\ref{map}).
\STATE Generate the index matrix ${\bf{K}}_z$ according to $\left\{ {{c_{{N_S}}},{c_{{N_S} - 1}}, \cdots ,{c_1}} \right\}$ as follows: for $a = 1, \cdots ,{N_S}$, we have $k_z^{n,j} = 1$ if and only if $n = \left\lfloor {\frac{{{c_a}{\rm{ + 1}}}}{T}} \right\rfloor$ and $j = \left\lfloor {\frac{{{c_a}{\rm{ + 1}}}}{N_t}} \right\rfloor$, while the remaining elements in matrix ${{\bf{K}}_z}$ are set to zero.\\
\STATE Let ${{\bf{U}}_z} \in {{\cal{R}}^{{N_t} \times 1}}$ be a vector denoting the cumulative activation numbers of LEDs corresponding to the $z$-th activation pattern, where $u_z^n = \sum\limits_{j = 1}^T {k_z^{n,j}} ,n = 1,2, \cdots {N_t}$ denotes the cumulative activation times of the $n$-th LED in the $k$-th active pattern ${\bf{K}}_z$.
\ENDFOR
\STATE Let the set of all possible active patterns be $\mathbb{K} =  \left\{ {{{\bf{K}}_1},{{\bf{K}}_2}, \ldots ,{{\bf{K}}_z}} \right\}$, and the corresponding set of actived times of each LED in each activation pattern be $\mathbb{U} = \left\{ {{{\bf{U}}_1},{{\bf{U}}_2}, \ldots ,{{\bf{U}}_z}} \right\}$.
\FOR {$z$ = 0 $ \to $ ${\rm{C}}({N_t}T,{N_S})-1$}
\STATE $n = 1$.
\WHILE {$n \le {N_t}$}
\IF {$u_z^n \ne 0$}
\STATE  Put ${{\bf{K}}_z}$ into set ${\mathbb{E}_n}$, and put ${{\bf{U}}_z}$ into set ${\mathbb{V}_n}$.
\STATE \textbf{break}
\ENDIF
\STATE $n = n + 1$.
\ENDWHILE
\ENDFOR
\FOR {$n$ = 1 $ \to $ ${N_t}$}
\FOR {$k$ = 1 $ \to $ ${\rm Card}\left( {\mathbb{V}_n} \right)$}
\STATE ${\bf{V}}(n) = \sum\limits_{i = 1}^k {{\mathbb{V}_n}\{ i\} }$.
\IF  {${{\bf{V}}\left( n \right) }  < \left\lceil {\frac{{{2^{{p_1}}}{N_S}}}{{{N_t}}}} \right\rceil$}
\STATE Put ${{\mathbb{E}_n}\{ k\} }$ into set ${\mathbb{O}}$, and put ${{\mathbb{V}_n}\{ k\} }$ into set ${\mathbb{W}}$.
\ENDIF
\ENDFOR
\ENDFOR
\WHILE {${\rm Card}\left({\mathbb{O}}\right) < {2^{{p_1}}}$}
\STATE ${\bf{W}} = \sum\limits_{i = 1}^{{\rm{Card}}\left( \mathbb{W} \right)} {\mathbb{W}\{ i\} } $. Note that ${\bf{W}}\left( n \right)$ is the $n$-th element of ${\bf{W}}$, which is equal to the number of activated times of the $n$-th LED.
\STATE $n* = \mathop {\min }\limits_n {\bf{W}}\left( n \right)$.
\STATE Solve ${{\bf{U}}_p} = \mathop {\arg \max }\limits_{{{\bf{U}}_p} \in {\mathbb K} \cap \overline {\mathbb O} } u_p^{n*}$. Put ${{\bf{K}}_p}$ into ${\mathbb{O}}$, and put ${{\bf{U}}_p}$ into ${\mathbb{W}}$.
\ENDWHILE
\WHILE {${\rm Card}\left( {\mathbb{O}} \right) > {2^{{p_1}}}$}
\STATE ${\bf{W}} = \sum\limits_{i = 1}^{{\rm{Card}}\left( \mathbb{W} \right)} {\mathbb{W}\{ i\} } $.
\STATE $n* = \mathop {\max }\limits_n {\bf{W}}\left( n \right)$.
\STATE Solve ${{\bf{U}}_p} = \mathop {\arg \max }\limits_{{{\bf{U}}_p} \in {\mathbb K} \cap \overline {\mathbb O} } u_p^{n*}$. Remove ${{\bf{K}}_p}$ from ${\mathbb{O}}$, and remove ${{\bf{U}}_p}$ from ${\mathbb{W}}$.
\ENDWHILE
\STATE \textbf{Output:} ${\mathbb{O}}$.
\end{spacing}
\end{algorithmic}
\label{algorithm1}
\end{breakablealgorithm}

\emph{Complexity Analysis}: If we use ES algorithm to solve problem (\ref{ilu_problem}), the corresponding search complexity can be expressed as ${\cal{O}} \left({\rm{C}}\left( { {\rm{Card}}\left( {\mathbb{K}} \right) , {\rm{Card}}\left( {\mathbb{O}} \right) } \right)\right)$. In contrast, the complexity of the proposed incremental algorithm is ${\cal{O}} \left( {{\rm C} \left( N_t T,N_S  \right) N_t} \right)$. It is obvious that the incremental algorithm has much lower complexity when compared with ES algorithm. We will further compare the illumination distribution of the sequential selection algorithm with that of the incremental algorithm in Section \ref{Results}.

\vspace{-0cm}
\subsection{Optimal Signal Form Of GDC}
\label{sec:siganl}
In the proposed GDC scheme, the dimming control is achieved by simultaneously adjusting the number of active elements in space-time blocks and the value of the forward current $I$. Note that $I$ is equal to the expected amplitude of the transmitted symbols as follows
\begin{eqnarray}
\label{average}
I = \frac{1}{M}\sum\limits_{m = {\rm{1}}}^M {{s_m}}.
\end{eqnarray}
Moreover, the overall normalized dimming level of GDC can be expressed as
\begin{eqnarray}\label{dimminglevel}
{\eta} = \frac{{{N_S}}}{{{N_t}T}}{\eta _e} \times 100\%
\end{eqnarray}
where $0 < {\eta _{\rm{e}}} \le 1$ is the optical power of each active element, and there is a linear relationship between the forward current $I$ and ${\eta _{\rm{e}}}$
\begin{eqnarray}
\label{etae}
{\eta _e} = \frac{{I - {I_{\rm{L}}}}}{{{I_{\rm{H}}} - {I_{\rm{L}}}}}.
\end{eqnarray}
Substituting (\ref{average}) into (\ref{etae}), we have
\begin{eqnarray}
\label{flr}
\sum\limits_{m = {\rm{1}}}^M {{s_m}} = M\left[ {{\eta _e}({I_{\rm{H}}} - {I_{\rm{L}}}) + {I_{\rm{L}}}} \right].
\end{eqnarray}
It can be observed from (\ref{signalform}) that the sequence $\left\{ {{s_1},{s_2}, \cdots ,{s_M}} \right\}$ is an equally spaced sequence. Therefore, for given ${\eta}$ and $N_S$, the optimal signal form of $s_m$ for minimizing BER can be formulated as
\begin{align}\label{abcdef}
&\mathop {\arg }\min \limits_{\lambda,{B_{\rm{L}}}} \;\;\;{{\rm{P}}\left( {\left| {{{\widehat s}_m} - {s_m}} \right| > \frac{\lambda }{2}\left| {{{\widehat {\bf{K}}}_z}= {{\bf{K}}_z}} \right.} \right)}\\
&\rm{s.\;t.}\;\;\scalebox{1}{${s_m}=\lambda m + {B_{\rm{L}}}$}\tag{\theequation a}\\
&\scalebox{1}{$\;\;\;\;\;\;\;{I_{\rm{L}}} < {s_1},{s_2}, \cdots ,{s_M} \le {I_{\rm{H}}}$} \tag{\theequation b}\\
&\scalebox{1}{$\;\;\;\;\;\;\;M = {2^{P - {p_1}}}$} \tag{\theequation c}\\
&\scalebox{1}{$\;\;\;\;\;\;\;{\eta _e} = \frac{{\eta {N_t}T}}{{{N_S}}}{\rm{ = }}\frac{{I - {I_{\rm{L}}}}}{{{I_{\rm{H}}} - {I_{\rm{L}}}}}$} \tag{\theequation d}\\
&\scalebox{1}{$\;\;\;\;\;\;\;\sum\limits_{m = {\rm{1}}}^M {{s_m}} = MI$} \tag{\theequation e}
\end{align}
where ${\rm{P}}( \cdot )$ denotes the probability function.

\vspace{0.2cm}
\noindent\textbf{$Lemma\ 1.$} The optimal values of $\lambda$ and $B_{\rm L}$ of (\ref{abcdef}) are given by
\begin{eqnarray}\label{oplam}
\begin{array}{l}
\lambda {\rm{ = }}\left\{ \begin{array}{l}
\frac{{{\rm{2}}\left( {{I_{\rm{H}}} - {I_{\rm{L}}}} \right){\eta _e}}}{{M - 1}},\;\;\;\;\;\;\;\;{\rm{if}}\;\;\;I \le \frac{{{I_{\rm{H}}} - {I_{\rm{L}}}}}{2}\\
\frac{{2(1 - {\eta _e})({I_{\rm{H}}} - {I_{\rm{L}}})}}{{M - 1}},\;\;\;{\rm{if}}\;\;I > \frac{{{I_{\rm{H}}} - {I_{\rm{L}}}}}{2}
\end{array} \right.\\
\;\;\;\;\;\;\;
\end{array}
\end{eqnarray}
and
\begin{eqnarray}\label{opbl}
\begin{array}{l}
{B_L}{\rm{ = }}\left\{ \begin{array}{l}
\frac{{\left( {2{\eta _e} - 1 - M} \right){I_{\rm{L}}} - 2{\eta _e}{I_{\rm{H}}}}}{{M - 1}},\;\;\;\;\;\;\;\;\;\;\;\;\;{\rm{if}}\;\;\;I \le \frac{{{I_{\rm{H}}} - {I_{\rm{L}}}}}{2}\\
\frac{{{\rm{2}}{I_{\rm{L}}}M(1 - {\eta _e}) - {I_{\rm{H}}}(1 + M - 2M{\eta _e})}}{{M - 1}},\;\;\;{\rm{if}}\;\;I > \frac{{{I_{\rm{H}}} - {I_{\rm{L}}}}}{2}
\end{array} \right.\\
\;\;\;\;\;\;\;
\end{array}
\end{eqnarray}
\begin{proof}
Please see Appendix A.
\end{proof}
Substituting (\ref{oplam}) and (\ref{opbl}) into (\ref{signalform}), we obtain the optimal PAM signal form that minimizes BER for GDC as
\begin{eqnarray}\label{PAM}
\begin{split}
s_m =
\left\{ \begin{split}
&\frac{{{\rm{2}}\left( {{I_{\rm{H}}} - {I_{\rm{L}}}} \right){\eta _e}}}{{M - 1}}m + \frac{{\left( {2{\eta _e} - 1 - M} \right){I_{\rm{L}}} - 2{\eta _e}{I_{\rm{H}}}}}{{M - 1}},\;\;\;\;\;\;\;\; \;\;\;\;\;\; \;\;\;\;\;\; \;\;\;\;\;{\rm{if}}\;\;\;I \le \frac{{{I_{\rm{H}}} - {I_{\rm{L}}}}}{2}\\
&\frac{{2(1 - {\eta _e})({I_{\rm{H}}} - {I_{\rm{L}}})}}{{M - 1}}m + \frac{{{\rm{2}}{I_{\rm{L}}}M(1 - {\eta _e}) - {I_{\rm{H}}}(1 + M - 2M{\eta _e})}}{{M - 1}}, \;\;{\rm{if}}\;\;\;I > \frac{{{I_{\rm{H}}} - {I_{\rm{L}}}}}{2}.
\end{split} \right.\;
\end{split}
\end{eqnarray}
We can observe that the optimal PAM signal form is a piece-wise function in terms of the forward current $I$.

\vspace{-0cm}
\section{GDC With Dimming Control Pattern Selection}
\label{sec:Signal Analysis of IM-ATD}
In Section III, we obtain the set of activation patterns and the optimal signal form of $s_m$ for a given $N_S$ under certain illumination constraint. However, since both the value of $N_S$ and the signal form can be adjusted with the illumination constraint, there may have multiple eligible $N_S$ for GDC that satisfy the illumination requirement.
In particular, with a given ${\eta}$, $N_S$ ranging from $\left\lceil {\eta {N_t}T} \right\rceil$ to ${{N_t}T}$ can satisfy the optical power constraint.
Hence, the configuration of GDC can be further optimized under the illumination constraints.
In this section, we first derive the BER upper bound of GDC, and GDC scheme with the minimum BER criterion (GDC-MBER) is proposed to obtain the optimal set of activation patterns by exhaustively searching all pairwise error and calculating the upper bound of theoretical BER. Moreover, two efficient low-complexity GDC schemes with the maximum free distance criterion (GDC-MFD) are proposed to reduce the computation complexity.

\vspace{-0cm}
\subsection{GDC-MBER}
The CPEP of GDC ${\rm{P}}({\bf{S}} \to {\bf{E}}\left| {\bf{H}} \right.)$ can be calculated as
\begin{eqnarray}
\begin{split}
&{\rm{P}}({\bf{S}} \to {\bf{E}}\left| {\bf{H}} \right.)\\
&= {\rm{P}}({\left\| {{\bf{Y}} - {\bf{H}}{\bf{S}}} \right\|_F^2} > {\left\| {{\bf{Y}} - {\bf{H}}{\bf{E}}} \right\|_F^2})\\
&= {\rm{P}} \left( {\left\| {{\bf{H}}{\bf{S}}} \right\|_F^2} - {\left\| {{\bf{H}}{\bf{E}}} \right\|_F^2} - 2{\left\| {{{\bf{Y}}^{\rm{T}}}{\bf{H}}\left( {{\bf{S}} - {\bf{E}}} \right)} \right\|_F} > 0 \right)\\
&= {\rm{P}}({\left\| {{\bf{H}}({\bf{S}} - {\bf{E}})} \right\|_F^2} - 2{\left\| {{{\bf{W}}^{\rm{T}}}{\bf{H}}\left( {{\bf{S}} - {\bf{E}}} \right)} \right\|_F} > 0)\\
&= {\rm{P}}(D > 0)
\end{split}
\end{eqnarray}
where the receiver detects the transmitted matrix ${\bf{S}}$ as ${\bf{E}} = {{\bf{K}}_w}{s_n}\left({\bf{E}} \ne {\bf{S}} \right)$ in error. In addition, $D \sim {\rm{{\cal N}}}({\alpha_D},\sigma _D^2)$ with ${\alpha_D} =  - {\left\| {{\bf{H}}({\bf{S}} - {\bf{E}})} \right\|_F^2}$ and $\sigma _D^2 = 2{N_0}{\left\| {{\bf{H}}({\bf{S}} - {\bf{E}})} \right\|_F^2}$ \cite{proakis1998digital}. Furthermore, we have
\begin{eqnarray}\label{medoptimal}
{\rm{P}}({\bf{S}} \to {\bf{E}}\left| {\bf{H}} \right.) = Q\left( {\sqrt {\frac{{{{\left\| {{\bf{H}}({\bf{S}} - {\bf{E}})} \right\|}_F^2}}}{{2{N_0}}}} } \right).
\end{eqnarray}
With the obtained CPEP, the BER of GDC can be calculated by the asymptotically tight union upper bound $P_b^{{\rm{upper}}}$ as \cite{ko2014tight,goldsmith2005wireless}
\begin{eqnarray}\label{unionbound}
P_b^{{\rm{upper}}} = \frac{1}{{2^{{P}}}{P}}\sum\limits_{{\bf{S}}} {\sum\limits_{{\bf{E}},{\bf{E}} \ne {\bf{S}}} {{\rm{P}}({\bf{S}} \to {\bf{E}}\left| {\bf{H}} \right.)n({\bf{S}},{\bf{E}})} }
\end{eqnarray}
\noindent where ${n({\bf{S}},{\bf{E}})}$ is the number of error bits in the pair ${\bf{S}}$ and ${\bf{E}}$.

Under the constraints of optical power and spectral efficiency, the activation pattern selection based on the minimum BER criterion can be obtained by
\begin{align}\label{opt-problem}
&\arg \mathop {\min }\limits_{{N_S}} \;{P_b^{{\rm{upper}}}}\\
&\rm{s.\;t.}\;\;\scalebox{1}{${\eta _e} = \frac{{\eta {N_t}T}}{{{N_S}}}$}\tag{\theequation a}\\
&\scalebox{1}{$\;\;\;\;\;\;\;M = {2^{(P - {p_1})}}.$} \tag{\theequation b}
\end{align}

Problem (\ref{opt-problem}) can be solved by exhaustively searching all possible pairwise errors ${\bf{S}}$,  ${\bf{E}}$ and $N_S$. However, since the exhaustive search requires high computation efforts, we further propose two simple suboptimal approaches based on the maximum free distance criterion, and their complexities will be compared in Section \ref{CA2}.
\vspace{-0cm}
\subsection{GDC-MFD1}
%$P_b$ can be approximated to (\textcolor{red}{cite}):

%, and ${d_{{\rm{free}}}^{{N_S}}}$ is the free distance of the received vector ${d_{{\rm{free}}}^{{N_S}}}$. , the approximation in (\ref{aproxBER}) can be used for BER analysis of GDC. According the nearest neighbor approximation defined in \cite{goldsmith2005wireless}, we can observe that $P_b$ has a close relationship with the free distance of the received vector ${d_{{\rm{free}}}^{{N_S}}}$
For high SNRs, the tight union upper bound $P_b^{{\rm{upper}}}$ is a function of the free distance of the received vector ${d_{{\rm{free}}}^{{N_S}}}$ \cite{minimumed,dmincite}, which can be expressed as \cite{TransmitPrecoded}
\begin{eqnarray}\label{aproxBER}
P_b^{{\rm{upper}}} = \chi Q\left( {\sqrt {\frac{{d_{{\rm{free}}}^{{N_S}}}}{{2{N_0}}}} } \right){n_{{d_{{\rm{free}}}}}}
\end{eqnarray}
where $\chi$ is the number of neighbor point \cite{TransmitPrecoded} and ${n_{{d_{{\rm{free}}}}}}$ denotes the number of error bits for the error events having the free distance
and ${d_{{\rm{free}}}^{{N_S}}}$ is defined as \cite{r1}
\begin{eqnarray}\label{diminimum}
{d_{{\rm{free}}}^{{N_S}}} = \mathop {\min }\limits_{{\bf{E}},{\bf{S}},{\bf{S}} \ne {\bf{E}}} {\left\| {{\bf{H}}\left( {{\bf{S}} - {\bf{E}}} \right)} \right\|_F^2} = \mathop {\min }\limits_{z,w,m,n} {\left\| {{\bf{H}}\left( {{{\bf{K}}_z}{s_m} - {{\bf{K}}_w}{s_n}} \right)} \right\|_F^2}.
\end{eqnarray}

Since VLC often has high SNRs \cite{highSNR,highSNRle} and ${P_b}$ decreases with ${d_{{\rm{free}}}^{{N_S}}}$, we can simplify the problem (\ref{opt-problem}) as
\begin{align}\label{subopt-problem}
&{\mathop {\arg } \max \limits_{{N_S}} \;\;{d_{{\rm{free}}}^{{N_S}}}}\\
&\rm{s.\;t.}\;\;\scalebox{1}{${\eta _e} = \frac{{\eta {N_t}T}}{{{N_S}}}$}\tag{\theequation a}\\
&\scalebox{1}{$\;\;\;\;\;\;M = {2^{(P - {p_1})}}$}. \tag{\theequation b}
\end{align}

In addition, for a given $N_S$, the free distance is derived from a pair of ${\bf{E}}$ and ${\bf{S}}$ with the smallest Euclidean distance, which can be classified into the following three cases:

\textbf{Case 1}: $\bf{S}$ and $\bf{E}$ differ only in the activation patterns. Therefore, the corresponding free distance ${d_{1,{{\rm{free}}} }^{{N_S}}}$ can be expressed as
\begin{eqnarray}
\begin{split}
{d_{1,{{\rm{free}}} }^{{N_S}}} = {\min _{m,z,w,z \ne w}}{\left\| {{{\bf{H}} }\left( {{{\bf{K}}_z}  - {{\bf{K}}_w} } \right){s_m}} \right\|_F^2}.
\end{split}
\end{eqnarray}

\textbf{Case 2}: $\bf{S}$ and $\bf{E}$ differ only in the transmitted symbols. The corresponding free distance ${d_{2,{{\rm{free}}} }^{{N_S}}}$ can be expressed as
\begin{eqnarray}
\begin{split}
{d_{2,{{\rm{free}}} }^{{N_S}}} = \mathop {\min _{z,m,n,m \ne n}} \left\| {{{\bf{H}} }{{\bf{K}}_z}\left( {{s_m} - {s_n}} \right)} \right\|_F^2.
\end{split}
\end{eqnarray}

\textbf{Case 3}: $\bf{S}$ and $\bf{E}$ differ both in the activation patterns and the transmitted symbols. The corresponding free distance ${d_{3,{{\rm{free}}} }^{{N_S}}}$ can be expressed as
\begin{eqnarray}\label{eq32}
\begin{split}
{d_{3,{{\rm{free}}} }^{{N_S}}} = &\mathop {\min }\limits_{\scriptstyle m,n,z,w\hfill\atop
\scriptstyle m \ne n,z \ne w\hfill} \left\| {{\bf{H}}\left( {{{\bf{K}}_z}{s_m} - {{\bf{K}}_w}{s_n}} \right)} \right\|_F^2\\
 = & \mathop {\min }\limits_{\scriptstyle m,n,z,w\hfill\atop
\scriptstyle m \ne n,z \ne w\hfill} \left\| {{\bf{H}}\left[ {{{\bf{K}}_z}\left( {{s_m} - {s_n}} \right) + \left( {{{\bf{K}}_z} - {{\bf{K}}_w}} \right){s_n}} \right]} \right\|_F^2\\
\ge & \sqrt {{{\left( {d_{1,{\rm{free}}}^{{N_S}}} \right)}^2} + {{\left( {d_{2,{\rm{free}}}^{{N_S}}} \right)}^2}}.
\end{split}
\end{eqnarray}

Therefore, ${d_{{\rm{free}}}^{{N_S}}}$ can be expressed as \cite{ko2014tight}
\begin{eqnarray}
{d_{{\rm{free}}}^{{N_S}}} = \min \left\{ {{d_{1,{{\rm{free}}} }^{{N_S}}},{d_{2,{{\rm{free}}} }^{{N_S}}},{d_{3,{{\rm{free}}} }^{{N_S}}}} \right\}.
\end{eqnarray}
From (\ref{eq32}), we can observe that ${d_{3,{{\rm{free}}} }^{{N_S}}}$ is always larger than ${d_{1,{{\rm{free}}} }^{{N_S}}}$ and ${d_{2,{{\rm{free}}} }^{{N_S}}}$, and thus the free distance ${d_{{{\rm{free}}} }^{{N_S}}}$ can be simplified as
\begin{eqnarray}\label{1231}
\begin{split}
{d_{{{\rm{free}}} }^{{N_S}}} = &\min \left\{ {{d_{1,{{\rm{free}}} }^{{N_S}}},{d_{2,{{\rm{free}}} }^{{N_S}}}} \right\}.
\end{split}
\end{eqnarray}

However, the computations of ${d_{1,{{\rm{free}}} }^{{N_S}}}$ and ${d_{2,{{\rm{free}}} }^{{N_S}}}$ still require exhaustive search over all active patterns and constellation points. When considering a large number of activation patterns and constellation size, exhaustively searching all possible pairs of ${\bf{S}}$ and ${\bf{E}}$ requires prohibitive computation complexity. To simplify the calculation, a lower bound of ${d_{{{\rm{free}}} }}$ is derived based on Rayleigh-Ritz theorem \cite{AntennaSelectionSM}. In particular, the lower bound of ${d_{1,{{\rm{free}}}}^{{N_S}}}$ can be derived as
\begin{eqnarray}\label{d1min1}
\begin{split}
{d_{1,{{\rm{free}}}}^{{N_S}}} \ge& {\mathop {\min }\limits_{z,w,z \ne w}}\Lambda _{\min ,({{\bf{HJ}}_{z,w}})}^2\mathop {\min }\limits_m {\left| {{s_m}} \right|^2} \\
=&{\mathop {\min }\limits_{z,w,z \ne w}} \Lambda _{\min ,({{\bf{HJ}}_{z,w}})}^2{\left| {{s_1}} \right|^2}
\end{split}
\end{eqnarray}
where ${{\bf{J}}_{z,w}} = {{\bf{K}}_z} - {{\bf{K}}_w}$, and $\Lambda _{\min ,({{\bf{HJ}}_{z,w}})}^2$ is the smallest eigenvalue of $ {{\bf{H}}{{\bf{J}}_{z,w}}} $.
On the other hand, the lower bound of ${d_{2,{{\rm{free}}}}^{{N_S}}}$ can be derived as
\begin{eqnarray}\label{d1min2}
\begin{split}
{d_{2,{{\rm{free}}} }^{{N_S}}} \ge &\mathop {\min_{z} } \left\| {{{\bf{H}} }{{\bf{K}}_z}} \right\|_F^2\mathop {\min_{m,n,m \ne n}} \left| {{s_m} - {s_n}} \right|^2\\
= &\mathop {\min_{z} } \Lambda _{\min ,({{\bf H{K}}_z})}^2{\left| \lambda  \right|^2}
\end{split}
\end{eqnarray}
where $\Lambda _{\min ,({{\bf H{K}}_z})}^2$ is the smallest eigenvalue of $ {{\bf{H}}{{\bf{K}}_z}} $.
Hence, the lower bound of the free distance ${d_{{{\rm{free}}} }^{{N_S}}}$ can be expressed as
\begin{eqnarray}\label{1231}
\begin{split}
{d_{{{\rm{free}}} }^{{N_S}}} = &\min \left\{ {{d_{1,{{\rm{free}}} }^{{N_S}}},{d_{2,{{\rm{free}}} }^{{N_S}}}} \right\}\\
\ge &\min \left\{ { {\mathop {\min }\limits_{z,w,z \ne w}} \Lambda _{\min ,({{\bf H{J}}_{z,w}})}^2{\left| {{s_1}} \right|^2},\mathop {\min }\limits_z \Lambda _{\min ,({{\bf H{K}}_z})}^2{\left| \lambda  \right|^2}} \right\}.
\end{split}
\end{eqnarray}

From the lower bound of ${d_{{{\rm{free}}} }^{{N_S}}}$ in (\ref{1231}), we can observe that only the minimum singular value of each $ {{\bf{H}}{{\bf{J}}_{z,w}}} $ and $ {{\bf{H}}{{\bf{K}}_z}} $ needs to be calculated and the constellation points no longer affect the lower bound of ${d_{{{\rm{free}}} }^{{N_S}}}$. Therefore, the required computational complexity can be reduced dramatically. In GDC-MFD1, we replace the objective function in (\ref{subopt-problem}) by the lower bound in (\ref{1231}) for achieving lower complexity.
The detail of GDC-MFD1 is summarized in \textbf{Algorithm} \ref{algorithm2}.
\begin{algorithm}
\small
\caption{GDC-MFD1.}
\begin{algorithmic}[1]
\STATE \textbf{Input:} $N_t,\;T,\;{\eta},\;P$.
\FOR {$N_S$ = $\left\lceil {\eta {N_t}} \right\rceil$ $ \to $ $N_t$}
\STATE Compute ${p_1} = \left\lfloor {{{\log }_2}({\rm{C}}({N_t}T,{N_S}))} \right\rfloor$, and then choose the minimal integer $M$ that satisfies $M \ge {2^{(P - {p_2})}}$.
\STATE Compute ${{{\left| {{s_1}} \right|}^2}}$ and ${{{\left| \lambda  \right|}^2}}$ according (\ref{dimminglevel}) and (\ref{PAM}).
\FOR {$z$ = $0$ $ \to $ ${2^{{p_1}}} - 1$}
\FOR {$w$ = $0$ $ \to $ $z - 1$}
\STATE Compute each $\mathop {\min }\limits_{z,w\hfill\atop
z \ne w\hfill} \Lambda _{\min ,({{\bf H {J}}_{z,w}})}^2$.
\ENDFOR
\STATE Compute each $\mathop {\min }\limits_z \Lambda _{\min ,({{\bf H {K}}_z})}^2$.
\ENDFOR
\STATE
Let ${d_{2,{\rm{free}}}^{{\rm{low}}}} = \mathop {\min }\limits_z \Lambda _{\min ,({{\bf H {K}}_z})}^2{\left| \lambda  \right|^2}$, ${d_{1,{\rm{free}}}^{{\rm{low}}}} = \mathop {\min }\limits_{z,w\hfill\atop
z \ne w\hfill} \Lambda _{\min ,({{\bf H {J}}_{z,w}})}^2{\left| {{s_1}} \right|^2}$.
$d_{{\rm{free}}}^{{N_S}} = \min \left\{ {d_{1,{\rm{free}}}^{{\rm{low}}},d_{2,{\rm{free}}}^{{\rm{low}}}} \right\}$.
\ENDFOR
\STATE
{\bf{Solve}} ${N_S} = \arg \mathop {\max }\limits_{{N_S}} d_{{\rm{free}}}^{{N_S}}$.
\end{algorithmic}
\label{algorithm2}
\end{algorithm}

\subsection{GDC-MFD2}
As discussed in Section IV.A, the high computational complexity of GDC-MBER is mainly due to the search of all pairs of ${\bf{S}}$ and ${\bf{E}}$, while GDC-MFD1 reduces computational complexity by simplifying CPEP into calculating the free distance, which can reduce the search range from $2^{2P}$ to $2^{2p_1}$.
However, when the space-time matrix has a large number of elements, the number of required floating point operations (FLOPs) in GDC-MFD1 is still large, which can result in potentially high computational complexity. To circumvent this problem, we further propose an algorithm named GDC-MFD2, which utilizes the time-invariance characteristics of the VLC channel to reduce the number of repetitive comparisons when calculating the free distance.

In GDC-MFD2, we assume ${\bf{J}}_{z,w}^{j} = \left[ {{\bf{K}}_z^{j} - {\bf{K}}_w^{j}} \right] \in {{\cal{R}}^{{N_t} \times 1}}$, where the superscript $j$ denotes the $j$-th column of a matrix, and thus we have
\begin{eqnarray}\label{dtamin1}
\begin{split}
{d_{{\rm{1,free}}}^{{N_S}}} &= \mathop {\min }\limits_{z,w,m} {\left\| {{\bf{H}}\left( {{{\bf{K}}_z} - {{\bf{K}}_w}} \right){s_m}} \right\|_F^2}\\
 &= \mathop {\min }\limits_{z,w} \sum\limits_{j = 1}^T {{{\left( {{\bf{J}}_{z,w}^{j}} \right)}^{\rm{T}}}{{\bf{H}}^{\rm{T}}}{\bf{HJ}}_{z,w}^{j}} \mathop {\min }\limits_m {\left| {{s_m}} \right|^2}\\
 &= \mathop {\min }\limits_{z,w} \sum\limits_{j = 1}^T {{{\left\| {{\bf{H}}{{\bf{J}}_{z,w}^{j}}} \right\|_F}^2}} {\left| {{s_1}} \right|^2}.
\end{split}
\end{eqnarray}
Since the values of the elements in ${\bf{J}}_{z,w}^{j}$ can only be 0, 1, and $-1$, the minimum value of $\sum\limits_{col = 1}^T {{{\left\| {{\bf{H}}{{\bf{J}}_{z,w}^{j}}} \right\|}_F^2}}$ is obtained when there is only one pair of non-zero elements in ${\bf{J}}_{z,w}^{j}$. In other words, in \textbf{Case 1}, ${d_{{\rm{1,free}}}}$ is obtained when the activation status of a pair of elements in the space-time matrix is detected incorrectly. Without loss of generality, we assume that the pair of error detected elements are located in the $row_1$-th row, the $col_1$-th column and the $row_2$-th row, the $col_2$-th column of the matrix ${{\bf{K}}_z}$. Besides, ${{\bf{H}}_J} \in {{\cal R}^{{2} \times {2}}}$ is the transmit matrix corresponding to the pair of error detected elements. Then, the free distance in (\ref{dtamin1}) corresponding to these two elements is defined as
\begin{eqnarray}\label{freedist}
F_{row_1,row_2}^{col_1,col_2} = {\bf{E}}{{\bf{H}}_J}{{\bf{E}}^{\rm{T}}}
\end{eqnarray}
where ${\bf{E}} \in \left\{ {\left( {1, - 1} \right),\left( { - 1,1} \right)} \right\}$ is the error detected vector.
For example, if ${{\bf{K}}_z} = {\left( {\begin{array}{*{20}{c}}
{\begin{array}{*{20}{c}}
1&0&0&1
\end{array}}\\
{\begin{array}{*{20}{c}}
1&1&0&0
\end{array}}
\end{array}} \right)^{\rm{T}}}$, ${{\bf{K}}_w} = {\left( {\begin{array}{*{20}{c}}
{\begin{array}{*{20}{c}}
1&0&1&1
\end{array}}\\
{\begin{array}{*{20}{c}}
0&1&0&0
\end{array}}
\end{array}} \right)^{\rm{T}}}$, and then ${{\bf{J}}_{z,w}} = {\left( {\begin{array}{*{20}{c}}
{\begin{array}{*{20}{c}}
0&0&{ - 1}&0
\end{array}}\\
{\begin{array}{*{20}{c}}
1&0&0&0
\end{array}}
\end{array}} \right)^{\rm{T}}}$.
With $col_1 = 1$, $row_1 = 1$, $col_2 = 2$ and $row_2 = 3$, we can further obtain the corresponding free distance $F_{1,3}^{1,2} = \left( {\begin{array}{*{20}{c}}
1&{ - 1}
\end{array}} \right){{\bf{H}}_J}{\left( {\begin{array}{*{20}{c}}
1&{ - 1}
\end{array}} \right)^T}$.

Due to the symmetry characteristic of the free distances in (\ref{freedist}) and the time-invariance characteristic of the VLC channel, we can further simplify the calculation of the free distance into the following two cases. For ease of illustration, we list all the free distances in Table \ref{tab:performance_comparison}.

First, for the diagonal of Table \ref{tab:performance_comparison}, we have $F_{row_1,row_2}^{col_1,col_1} = F_{row_1,row_2}^{col_2,col_2}$, which is marked in blue. This is because different columns represent different time slots and due to the time-invariance characteristics of VLC channel, ${{\bf{H}}_J}$ is the same for $F_{row_1,row_2}^{col_1,col_1}$ and $F_{row_1,row_2}^{col_2,col_2}$. As a result, the values of the elements in each blue square matrix in Table \ref{tab:performance_comparison} are the same, and thus we only need to calculate the elements in one of the blue square matrix in Table \ref{tab:performance_comparison}.

Second, we have $F_{row_1,row_2}^{col_1,col_2} = F_{row_2,row_1}^{col_1,col_2}$ and $F_{row_1,row_2}^{col_1,col_2} = F_{row_1,row_2}^{col_1,col_3}$ for $col_3 \ne col_1 \ne col_2$ due to a similar reason. That implies that Table \ref{tab:performance_comparison} is symmetric and each sub-matrix in pink is also symmetric. Therefore, only the elements of each pink lower triangular matrix needs to be calculated. In this way, the calculation complexity of the free distance is significantly reduced.

\definecolor{mypink}{rgb}{.99,.71,.95}
\definecolor{mycyan}{cmyk}{.3,0,0,0}

\renewcommand{\arraystretch}{1.5}
\begin{table}
  \centering
  \fontsize{6}{8}\selectfont
  \caption{Free distance table.}
  \label{tab:performance_comparison}
  \resizebox{\textwidth}{44mm}{
    \begin{tabular}{|c|c|c|c|c|c|c|c|c|c|c|c|c|c|c|c|}
    \hline
    \multirow{2}{*}{\;}&\multirow{2}{*}{}&
    \multicolumn{4}{c|}{${col_2 = 1}$}&\multicolumn{4}{c|}{${col_2 = 2}$}&\multicolumn{1}{c|}{$\cdots$}&\multicolumn{4}{c|}{${col_2 = T}$}\cr\cline{3-15}
    &&${row_2=1}$&${row_2=2}$&$\cdots$&${row_2={{N_t}}}$
    &${row_2=1}$&${row_2=2}$&$\cdots$&${row_2={{N_t}}}$&$\cdots$
    &${row_2=1}$&${row_2=2}$&$\cdots$&${row_2={{N_t}}}$\cr
    \hline

    \multirow{4}{*}{${col_1 = 1}$}&
    ${row_1=1}$&\cellcolor{mycyan}$\backslash$&\cellcolor{mycyan}{${F_{1,2}^{1,1}}$}&\cellcolor{mycyan}{$\cdots$}&\cellcolor{mycyan}{${F_{1,{N_t}}^{1,1}}$}
    &\cellcolor{mypink}${F_{1,1}^{1,2}}$&\cellcolor{mypink}${F_{1,2}^{1,2}}$&\cellcolor{mypink}$\cdots$&\cellcolor{mypink}${F_{1,{N_t}}^{1,2}}$
    &&\cellcolor{mypink}${F_{1,1}^{1,T}}$&\cellcolor{mypink}${F_{1,2}^{1,T}}$&\cellcolor{mypink}$\cdots$&\cellcolor{mypink}${F_{1,{N_t}}^{1,T}}$
    \cr\cline{2-15}

    &${row_1=2}$&\cellcolor{mycyan}{${F_{2,1}^{1,1}}$}&\cellcolor{mycyan}$\backslash$&\cellcolor{mycyan}{$\cdots$}&\cellcolor{mycyan}{${F_{2,{N_t}}^{1,1}}$}
    &\cellcolor{mypink}{${F_{2,1}^{1,2}}$}&\cellcolor{mypink}${F_{2,2}^{1,2}}$&\cellcolor{mypink}$\cdots$&\cellcolor{mypink}${F_{2,{N_t}}^{1,2}}$
    &&\cellcolor{mypink}{${F_{2,1}^{1,T}}$}&\cellcolor{mypink}${F_{2,2}^{1,T}}$&\cellcolor{mypink}$\cdots$&\cellcolor{mypink}${F_{2,{N_t}}^{1,T}}$
    \cr\cline{2-15}

    &$\vdots$&\cellcolor{mycyan}{$\vdots$}&\cellcolor{mycyan}{$\vdots$}&\cellcolor{mycyan}$\ddots$&\cellcolor{mycyan}{$\vdots$}
    &\cellcolor{mypink}{$\vdots$}&\cellcolor{mypink}{$\vdots$}&\cellcolor{mypink}$\ddots$&\cellcolor{mypink}$\vdots$
    &&\cellcolor{mypink}{$\vdots$}&\cellcolor{mypink}{$\vdots$}&\cellcolor{mypink}$\ddots$&\cellcolor{mypink}$\vdots$
    \cr\cline{2-15}

    &${row_1={{N_t}}}$&\cellcolor{mycyan}{${F_{{N_t},1}^{1,1}}$}&\cellcolor{mycyan}{${F_{{N_t},2}^{1,1}}$}&\cellcolor{mycyan}{$\cdots$}&\cellcolor{mycyan}$\backslash$
    &\cellcolor{mypink}{${F_{{N_t},1}^{1,2}}$}&\cellcolor{mypink}{${F_{{N_t},2}^{1,2}}$}&\cellcolor{mypink}{$\cdots$}&\cellcolor{mypink}${F_{{N_t},{N_t}}^{1,2}}$
    &&\cellcolor{mypink}{${F_{{N_t},1}^{1,T}}$}&\cellcolor{mypink}{${F_{{N_t},2}^{1,T}}$}&\cellcolor{mypink}{$\cdots$}&\cellcolor{mypink}${F_{{N_t},{N_t}}^{1,T}}$\cr
    \hline

    \multirow{4}{*}{${col_1 = 2}$}&
    ${row_1=1}$&\cellcolor{mypink}${F_{1,1}^{2,1}}$&\cellcolor{mypink}${F_{1,2}^{2,1}}$&\cellcolor{mypink}$\cdots$&\cellcolor{mypink}${F_{1,{N_t}}^{2,1}}$
    &\cellcolor{mycyan}$\backslash$&\cellcolor{mycyan}{${F_{1,2}^{2,2}}$}&\cellcolor{mycyan}{$\cdots$}&\cellcolor{mycyan}{${F_{1,{N_t}}^{2,2}}$}
    &&\cellcolor{mypink}${F_{1,1}^{2,T}}$&\cellcolor{mypink}${F_{1,2}^{2,T}}$&\cellcolor{mypink}$\cdots$&\cellcolor{mypink}${F_{1,{N_t}}^{2,T}}$
    \cr\cline{2-15}

    \;&${row_1=2}$&\cellcolor{mypink}{${F_{2,1}^{2,1}}$}&\cellcolor{mypink}${F_{2,2}^{2,1}}$&\cellcolor{mypink}$\cdots$&\cellcolor{mypink}${F_{2,{N_t}}^{2,1}}$
    &\cellcolor{mycyan}{${F_{2,1}^{2,2}}$}&\cellcolor{mycyan}$\backslash$&\cellcolor{mycyan}{$\cdots$}&\cellcolor{mycyan}{${F_{2,{N_t}}^{2,2}}$}
    &&\cellcolor{mypink}{${F_{2,1}^{2,T}}$}&\cellcolor{mypink}${F_{2,2}^{2,T}}$&\cellcolor{mypink}$\cdots$&\cellcolor{mypink}${F_{2,{N_t}}^{2,T}}$
    \cr\cline{2-15}

    \;&$\vdots$&\cellcolor{mypink}{$\vdots$}&\cellcolor{mypink}{$\vdots$}&\cellcolor{mypink}$\ddots$&\cellcolor{mypink}$\vdots$
    &\cellcolor{mycyan}{$\vdots$}&\cellcolor{mycyan}{$\vdots$}&\cellcolor{mycyan}$\ddots$&\cellcolor{mycyan}{$\vdots$}
    &&\cellcolor{mypink}{$\vdots$}&\cellcolor{mypink}{$\vdots$}&\cellcolor{mypink}$\ddots$&\cellcolor{mypink}$\vdots$
    \cr\cline{2-15}

    \;&${row_1={{N_t}}}$&\cellcolor{mypink}{${F_{{N_t},1}^{2,1}}$}&\cellcolor{mypink}{${F_{{N_t},2}^{2,1}}$}&\cellcolor{mypink}{$\cdots$}&\cellcolor{mypink}${F_{{N_t},{N_t}}^{2,1}}$
    &\cellcolor{mycyan}{${F_{{N_t},1}^{2,2}}$}&\cellcolor{mycyan}{${F_{{N_t},2}^{2,2}}$}&\cellcolor{mycyan}{$\cdots$}&\cellcolor{mycyan}$\backslash$
    &&\cellcolor{mypink}{${F_{{N_t},1}^{2,T}}$}&\cellcolor{mypink}{${F_{{N_t},2}^{2,T}}$}&\cellcolor{mypink}{$\cdots$}&\cellcolor{mypink}${F_{{N_t},{N_t}}^{2,T}}$\cr
    \hline

    \multirow{1}{*}{$\cdots$}&&&&&&&&&&&&&&\cr
    \hline

    \multirow{4}{*}{${col_1 = T}$}&
    ${row_1=1}$&\cellcolor{mypink}${F_{1,1}^{T,1}}$&\cellcolor{mypink}${F_{1,2}^{T,1}}$&\cellcolor{mypink}$\cdots$&\cellcolor{mypink}${F_{1,{N_t}}^{T,1}}$
    &\cellcolor{mypink}${F_{1,1}^{T,2}}$&\cellcolor{mypink}${F_{1,2}^{T,2}}$&\cellcolor{mypink}$\cdots$&\cellcolor{mypink}${F_{1,{N_t}}^{T,2}}$
    &&\cellcolor{mycyan}$\backslash$&\cellcolor{mycyan}{${F_{1,2}^{T,T}}$}&\cellcolor{mycyan}{$\cdots$}&\cellcolor{mycyan}{${F_{1,{N_t}}^{T,T}}$}
    \cr\cline{2-15}\;

    &${row_1=2}$&\cellcolor{mypink}{${F_{2,1}^{T,1}}$}&\cellcolor{mypink}${F_{2,2}^{T,1}}$&\cellcolor{mypink}$\cdots$&\cellcolor{mypink}${F_{2,{N_t}}^{T,1}}$
    &\cellcolor{mypink}{${F_{2,1}^{T,2}}$}&\cellcolor{mypink}${F_{2,2}^{T,2}}$&\cellcolor{mypink}$\cdots$&\cellcolor{mypink}${F_{2,{N_t}}^{T,2}}$
    &&\cellcolor{mycyan}{${F_{2,1}^{T,T}}$}&\cellcolor{mycyan}$\backslash$&\cellcolor{mycyan}{$\cdots$}&\cellcolor{mycyan}{${F_{2,{N_t}}^{T,T}}$}
    \cr\cline{2-15}\;

    &$\vdots$&\cellcolor{mypink}{$\vdots$}&\cellcolor{mypink}{$\vdots$}&\cellcolor{mypink}$\ddots$&\cellcolor{mypink}$\vdots$&\cellcolor{mypink}{$\vdots$}&\cellcolor{mypink}{$\vdots$}&\cellcolor{mypink}$\ddots$&\cellcolor{mypink}$\vdots$
    &&\cellcolor{mycyan}{$\vdots$}&\cellcolor{mycyan}{$\vdots$}&\cellcolor{mycyan}$\ddots$&\cellcolor{mycyan}{$\vdots$}
    \cr\cline{2-15}

    \;&${row_1={{N_t}}}$&\cellcolor{mypink}{${F_{{N_t},1}^{T,1}}$}&\cellcolor{mypink}{${F_{{N_t},2}^{T,1}}$}&\cellcolor{mypink}{$\cdots$}&\cellcolor{mypink}${F_{{N_t},{N_t}}^{T,1}}$
    &\cellcolor{mypink}{${F_{{N_t},1}^{T,2}}$}&\cellcolor{mypink}{${F_{{N_t},2}^{T,2}}$}&\cellcolor{mypink}{$\cdots$}&\cellcolor{mypink}${F_{{N_t},{N_t}}^{T,2}}$
    &&\cellcolor{mycyan}{${F_{{N_t},1}^{T,T}}$}&\cellcolor{mycyan}{${F_{{N_t},2}^{T,T}}$}&\cellcolor{mycyan}{$\cdots$}&\cellcolor{mycyan}$\backslash$\cr\hline
    \end{tabular}
}
\end{table}

The calculation of ${d_{{\rm{2,free}}}^{{N_S}}}$ in GDC-MFD2 can also be simplified to
\begin{eqnarray}\label{dtamin2}
\begin{split}
{d_{{\rm{2,free}}}^{{N_S}}} &= \mathop {\min }\limits_{z,n,m,m \ne n} {\left\| {{\bf{H}}{{\bf{K}}_z}\left( {{s_m} - {s_n}} \right)} \right\|_F^2} \\
&= \mathop {\min }\limits_z \sum\limits_{col = 1}^T {{{\left\| {{\bf{HK}}_z^{col}} \right\|}^2}} {\left| {{s_1}} \right|^2}\\
&= \mathop {\min }\limits_z \left|{{\bf{A}}_z}{{\bf{H}}_z}{\bf{A}}_z^{\rm{T}}\right| {\left| {{s_1}} \right|_F^2}
\end{split}
\end{eqnarray}
where ${\bf{K}}_z^{col}$ denotes the $col$-th column of matrix ${\bf{K}}_z$, ${{\bf{A}}_z} \in {{\cal{R}}^{1 \times {N_S}}}$ is a vector whose elements are all of $1$ and ${{\bf{H}}_z} \in {{\cal{R}}^{{N_S} \times {N_S}}}$ denotes the transmit matrix corresponding to the $z$-th activation pattern.
%Moreover, we define ${\bf{D}} \in {{\cal{R}}^{{N_t}T \times {N_t}T}}$ as a block diagonal matrix whose diagonal contains of $T$ repeated ${{{\bf{H}}^{\rm{T}}}{\bf{H}}}$, and several ${N_t} \times {N_t}$-dimensional ${\bf{0}}$ matrices elsewhere. We denote matrix ${\bf{D}}$ as:
%\begin{eqnarray}
%{\bf{D}} = {\rm{diag}}\underbrace {\left\{ {{{\bf{H}}^{\rm{T}}}{\bf{H}},{{\bf{H}}^{\rm{T}}}{\bf{H}}, \cdots ,{{\bf{H}}^{\rm{T}}}{\bf{H}}} \right\}}_T.
%\end{eqnarray}
Since both ${{\bf{H}}_J}$ and ${{\bf{H}}_z}$ are composed of the elements in ${{\bf{H}}}$ and $0 < {N_S} \le {N_t}T$, the number of required FLOPs for calculating ${d_{{\rm{1,free}}}^{{N_S}}}$ and ${d_{{\rm{2,free}}}^{{N_S}}}$ in GDC-MFD2 is smaller than that in GDC-MFD1. Then, we can obtain the free distance ${d_{{{\rm{free}}} }^{{N_S}}}$ according to (\ref{1231}).

\vspace{-0cm}
\subsection{Complexity Analysis}\label{CA2}
In this section, we analyze the computation complexities of GDC-MBER, GDC-MFD1 and GDC-MFD2. We express the computation complexity in terms of the complexity orders ${\mathcal{O}}\left(  \cdot  \right)$. Note that although FLOP counting cannot characterize the exact computation complexity, it is able to capture the order of the computation load and is sufficient for the complexity analysis \cite{complexity1}.

It is straightforward from (\ref{opt-problem}) that the complexity of GDC-MBER depends on the computation complexity of ${P_b^{{\rm{upper}}}}$, which requires to calculate all the possible CPEPs. For a given $N_S$, since the number of all the possible CPEPs is ${2^P}\left( {{2^P} - 1} \right)$, the overall computational complexity of GDC-MBER is
\begin{eqnarray}
\begin{split}
{\mathcal{O}_{MBER}} = \left( {{2^{2P}} - {2^P}} \right){\mathcal{O}_{P_b^{{\rm{upper}}}}}
\end{split}
\end{eqnarray}
where ${\mathcal{O}_{P_b^{{\rm{upper}}}}}$ is the number of required FLOPs for calculating a CPEP and it is given by
\begin{eqnarray}
\begin{split}
{\mathcal{O}_{P_b^{{\rm{upper}}}}} = &\underbrace {{N_t}T}_{{\rm{calculate }} \; {{\bf{K}}_z}{s_m}} + \underbrace {{N_t}T}_{{\rm{calculate }} \; {{\bf{K}}_w}{s_n}} + \underbrace {2{N_t}{N_R}T - {N_R}T + {N_t}T}_{{\rm{calculate  }} \; {\bf{H}}\left( {{\bf{S}} - {\bf{E}}} \right)} + \underbrace {{\rm{2}}N_R^2\left( {T - 1} \right)}_{{\rm{calculate }} \; \left\| {{\bf{H}}\left( {{\bf{S}} - {\bf{E}}} \right)} \right\|_F^2}.
\end{split}
\end{eqnarray}

In contrast, according to (\ref{subopt-problem}), we can observe that the complexities of both GDC-MFD1 and GDC-MFD2 depend on the computation complexity of ${{d_{{\rm{free}}}^{{N_S}}}}$. In GDC-MFD1, the number of all possible ${d_{{\rm{1,free}}}}$ is ${2^{{p_1}}}\left( {{2^{{p_1}}} - 1} \right)$ and the number of all possible ${d_{{\rm{2,free}}}}$ in (\ref{d1min2}) is ${2^{{p_1}}}$. Hence, for a given $N_S$, the computational complexity of solving (\ref{subopt-problem}) for GDC-MFD1 is:
\begin{eqnarray}
{\mathcal{O}_{MFD1}} = {2^{{p_1}}}\left( {{2^{{p_1}}} - 1} \right){\mathcal{O}_{{d_{{\rm{1,MFD1}}}}}} + {2^{{p_1}}}{\mathcal{O}_{{d_{{\rm{2,MFD1}}}}}}
\end{eqnarray}
where ${\mathcal{O}_{{d_{{\rm{1,MFD1}}}}}}$ and ${\mathcal{O}_{{d_{{\rm{2,MFD1}}}}}}$ are the number of FLOPs for calculating ${{d_{{\rm{1,free}}}^{{N_S}}}}$ and ${{d_{{\rm{2,free}}}^{{N_S}}}}$, respectively:
\begin{eqnarray}
{\mathcal{O}_{{d_{{\rm{1,MFD1}}}}}} = \underbrace {2{N_t}{N_R}T - {N_R}T + {N_t}T}_{{\rm{calculate}}\ {\bf{H}}\left( {{{\bf{K}}_z} - {{\bf{K}}_w}} \right)} + \underbrace {2{N_R}{T^2} + 13{T^2}}_{{\rm{SVD}}\;{\rm{matrix}}\;{\rm{decomposition\ for}}\;{\bf{H}}\left( {{{\bf{K}}_z} - {{\bf{K}}_w}} \right)}
\end{eqnarray}
and
\begin{eqnarray}
{\mathcal{O}_{{d_{{\rm{2,MFD1}}}}}} = \underbrace {2{N_t}{N_R}T - {N_R}T}_{{\rm{calculate  }}\ {\bf{H}}{{\bf{K}}_z}} + \underbrace {2{N_R}{T^2} + 13{T^2}}_{{\rm{SVD\ matrix\ decomposition\ {\rm{ for \ }}{\bf{H}}{{\bf{K}}_z}}}}.
\end{eqnarray}

In GDC-MFD2, we exploit the time invariance feature of the VLC channel to further reduce the computational complexity. As discussed in Section IV.C, since only ${{{N_t}}^2}$ independent values in Table \uppercase\expandafter{\romannumeral1} need to be calculated, the computational complexity of GDC-MFD2 can be calculated as:
\begin{eqnarray}
\begin{split}
{\mathcal{O}_{MFD2}} &= { {{N_t}} ^2}{\mathcal{O}_{{d_{{\rm{1,MFD2}}}}}} + {2^{{p_1}}}{\mathcal{O}_{{d_{{\rm{2,MFD2}}}}}}\\
&= { {{N_t}} ^2}\underbrace {{\rm{2}} \times \left( {{\rm{8 - 2}}} \right)}_{{\rm{calculate }}\ F_{row1,row2}^{col1,col2}}{\rm{ + }}{2^{{p_1}}}\underbrace {{\rm{2}} \times \left( {{\rm{2}}N_S^2{\rm{ - 2}}{N_S}} \right)}_{{\rm{calculate }}\ {{\bf{A}}_z}{{\bf{H}}_A}{\bf{A}}_z^{\rm{T}}}
\end{split}
\end{eqnarray}
where ${\mathcal{O}_{{d_{{\rm{1,MFD2}}}}}}$ and ${\mathcal{O}_{{d_{{\rm{2,MFD2}}}}}}$ are the number of required FLOPs for calculating ${{d_{{\rm{1,free}}}^{{N_S}}}}$ and ${{d_{{\rm{2,free}}}^{{N_S}}}}$ in GDC-MFD2, respectively.

In general, the search complexity of GDC-MBER is $\mathcal{O}({2^{2P}})$, while that of GDC-MFD1 and GDC-MFD2 are $\mathcal{O}({2^{{2p_1}}})$ and $\mathcal{O}({2^{{p_1}}})$, respectively. Note that $0 < {p_1} < 2{p_1} < 2({p_1} + {p_2}) = 2P$, and ${p_2}$ may be large for high transmission rate. Therefore, GDC-MFD2 has lower computational complexity than GDC-MBER and GDC-MFD1.
\vspace{-0.cm}
\section{Theoretical and Simulation Results}
\label{Results}
In this section, we evaluate the illumination and communication performance of the proposed GDC schemes. Here, the proposed GDC schemes include GDC-MBER, GDC-MFD1 and GDC-MFD2. Since the GDC schemes can be deemed as a general dimming control scheme that combines current AD, DD, and SD schemes, we adopt the HD scheme in \cite{HybridDimming} that combines AD and SD (denoted by ${\rm{H}}{{\rm{D}}_{{\rm{AD - SD}}}}$), and the DD scheme in \cite{PWMdimming1} as baseline schemes. Unless otherwise stated, the data rate of all the schemes are set to 8 bits/symbol and PAM modulation scheme is employed for all the scheme. The key system parameters are summarized in Table II.%\uppercase\expandafter{\romannumeral2}.
\begin{table}
\small
\vspace{-0.2cm}
\centering
\caption{Simulation Parameters.}
\setlength{\abovecaptionskip}{-0.2cm}
\begin{tabular}{ccc}
\toprule
Name of Parameters & Values \\
\hline
Turn-on current, ${{I_{\rm{L}}}}$ & 0.1 A \\
Maximum allowed current, ${{I_{\rm{H}}}}$ & 2 A\\
Number of LEDs, ${N_t}$ & 4\\
Number of PDs, ${N_R}$ & 4\\
Length of the space-time matrix in time domain, $T$ & 2\\
Semiangle at half power, ${\Phi _{{1 \mathord{\left/ {\vphantom {1 2}} \right.\kern-\nulldelimiterspace} 2}}}$ & ${\rm{60}}^\circ$\\
Receiver FOV semiangle, ${\Psi _{{1 \mathord{\left/{\vphantom {1 2}} \right.\kern-\nulldelimiterspace} 2}}}$ & ${\rm{40}}^\circ$\\
\multirow{2}{*}{Coordinates of the LEDs} & (1,1,2.5);(1,3,2.5)\\
 & (3,1,2.5);(3,3,2.5)\\
\multirow{2}{*}{Coordinates of the PDs} & (1.9,1.9,0.75);(1.9,2.1,0.75)\\
&(2.1,1.9,0.75);(2.1,2.1,0.75)\\
\toprule
\end{tabular}
\vspace{-0.5cm}
\end{table}

\subsection{Illumination Evaluations}
In illumination engineering, illumination uniformity is always an important metric for indoor illumination, which is typically measured by the uniformity illuminance ratio (UIR). In particular, UIR can be defined as $\frac{{{U_{\min }}}}{{{U_{{\rm{ave}}}}}}$, where ${U_{\min }}$ is the minimum illuminance and ${U_{{\rm{ave}}}}$ is the average illuminance at the receiver plane. Since the maximum UIR is different under different dimming levels, the normalized uniform illumination ratio (NUIR) is employed as the performance metric for comparisons. Specifically, we define the UIDR at a target dimming level $\eta$ as
\begin{eqnarray}
{\rm{NUIR}}\left( \eta  \right) = \frac{{{\rm{UIR}}\left( \eta  \right)}}{{{\rm{UIR}}{{\left( \eta  \right)}_{\rm non-spatial}}}}
\end{eqnarray}
where ${{\rm{UIR}}{{\left( \eta  \right)}_{\rm non-spatial}}}$ denotes the UIR of non-spatial dimming scheme\footnote{Here, the non-spatial dimming scheme means the dimming control scheme control does not include spatial dimming. In particular, the non-spatial dimming scheme include AD, DD and HD that combines AD and DD.} at a target dimming level $\eta$. Since for non-spatial dimming schemes, all the LEDs have the same activation probability, ${{\rm{UIR}}{{\left( \eta  \right)}_{\rm non-spatial}}}$ is employed as the normalization factor here.
%Note that since the optical power of each LED is the same at various dimming level, the UIR is a fixed constant in non-spatial dimming scheme, such as AD and DD. However, in GDC, the active elements of space-time matrices are different at different dimming levels and the cumulative activated times of each LED are different at various activation patterns. However, since the cumulative activated times for all LEDs in non-spatial dimming schemes are same at different dimming levels, the UIR of GDC with index mapping may be smaller than that of non-spatial dimming schemes.
%In this subsection, we evaluate the illumination uniformity of the proposed GDC scheme. In illumination engineering, the illumination uniformity if typically quantified by uniformity illumination ratio, which is defined as:
%For a given ${p_1}$, the algorithm proposed in Section III. B selects the set of activation patterns under the constraint of illumination uniformity. In contrast, the set of activation patterns obtained by sequentially selecting from the $0$-th activation patterns \cite{bacsar2013orthogonal} is not limited by the constraint of illumination uniformity, and thus it is not suitable for the scene with the limitation of uniform illumination.

\begin{figure}
\centering
\includegraphics[width=0.6\textwidth]{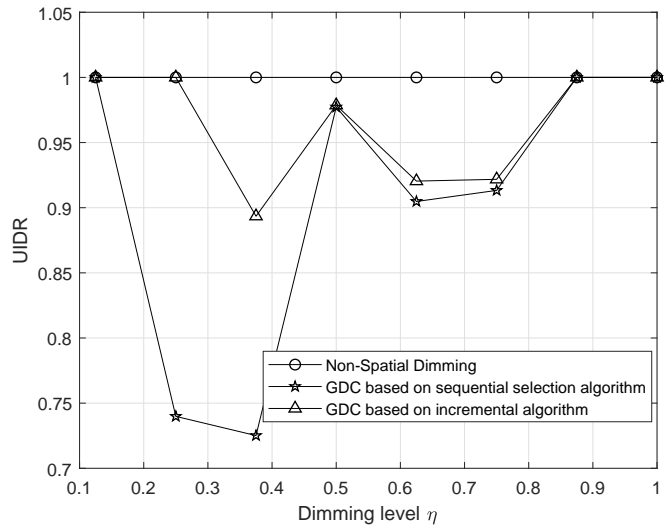}
\caption{UIDR comparison of the non-spatial dimming scheme, GDC based on the proposed incremental algorithm and GDC based on the sequential selection algorithm at various dimming levels.}
\label{fig:uir_dimming}
\end{figure}

%
To validate the efficiency of the proposed incremental algorithm for index mapping, Fig. 2 compares the UIDRs of non-spatial dimming scheme, GDC based on the incremental algorithm in \textbf{Algorithm 1} and GDC based on sequential selection algorithm used in \cite{bacsar2013orthogonal}. Note that since the ES algorithm is too complicated to implement, it is not illustrated in Fig. \ref{fig:uir_dimming}. For instance, when $N_S = 4$ and $N_tT = 8$, we have ${\rm{Card}}\left( {\mathbb{K}} \right) = {\rm{C}}\left( {8,4} \right) = 70$ and ${\rm{Card}}\left( {\mathbb{O}} \right) = {2^6} = 64$. Hence, the search complexity of ES can be calculated as ${\rm{C}}\left( { {\rm{Card}}\left( {\mathbb{K}} \right) , {\rm{Card}}\left( {\mathbb{O}} \right) } \right) = {\rm{C}}\left( {70,64} \right) \approx 1.3 \times {10^8}$, which is computationally prohibitive.
%Furthermore, the diagram of illumination distribution is shown in Fig. \ref{fig:comp} at the dimming level of 20\%.

Generally, the illuminance distribution in the room is more uniform when the value of UIDR is larger. It is shown in Fig. 2 that the UIDR of GDC based on the incremental algorithm is higher than that of GDC based on sequential selection algorithm. The minimum UIDR of GDC based on sequential selection is around 72.5\%, while the minimum UIDR of GDC based on the incremental algorithm is around 90\%, which indicates that the proposed GDC based on the incremental algorithm achieves better uniform illumination than that based on the sequential selection algorithm. For the non-spatial dimming control scheme, since the variation of the dimming level does not change the activation state of LEDs, its UIDR is always 1.

\begin{figure}[t]
\setlength{\abovecaptionskip}{0.2cm}   %调整图片标题与图距离
\setlength{\belowcaptionskip}{-8pt}   %调整图片标题与下文距离
 \centering
    \subfigure[GDC based on sequential selection.]
    {
    \includegraphics[width=0.47\linewidth]{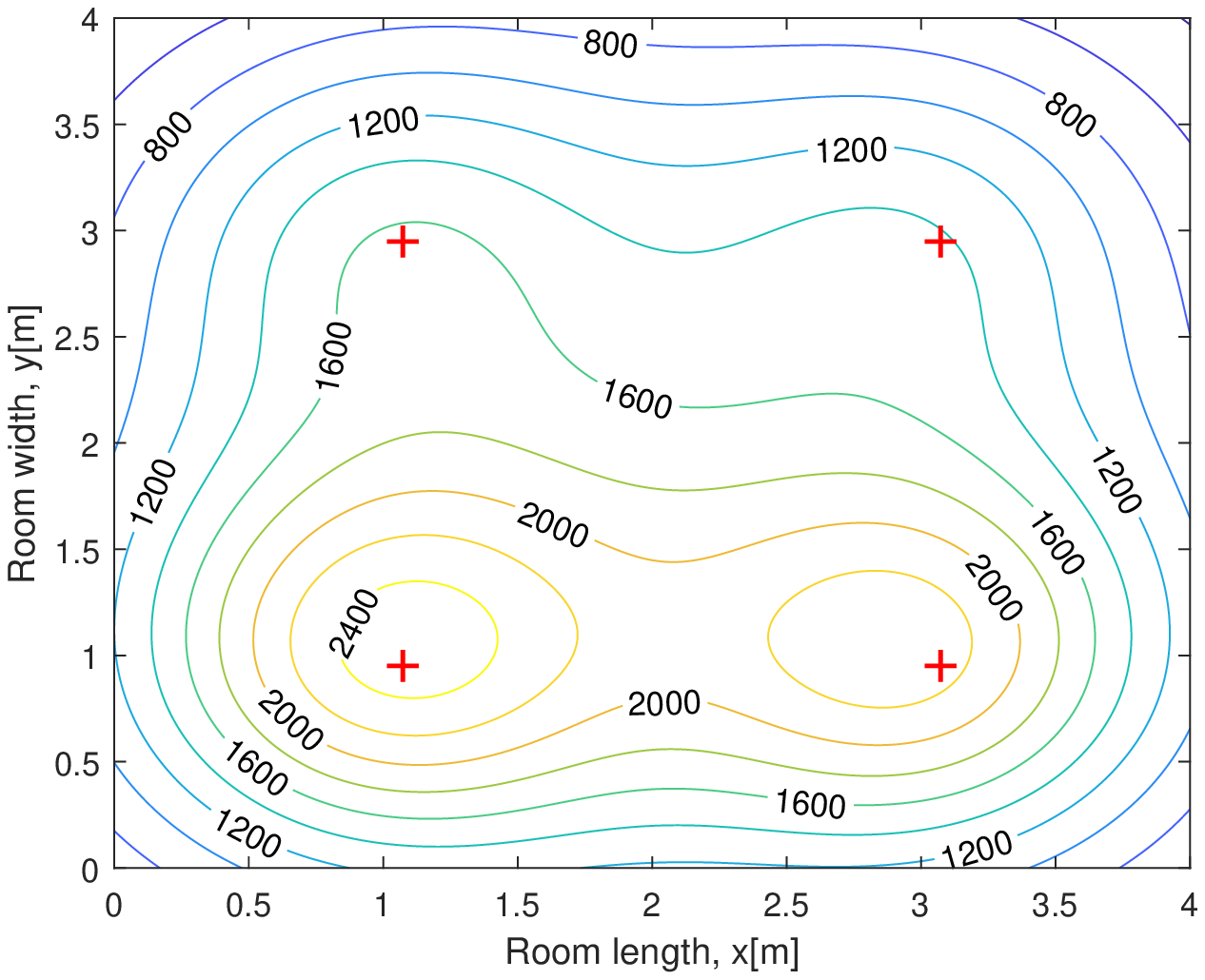}
    }
    \subfigure[GDC based on the proposed incremental algorithm.]
    {
    \includegraphics[width=0.47\linewidth]{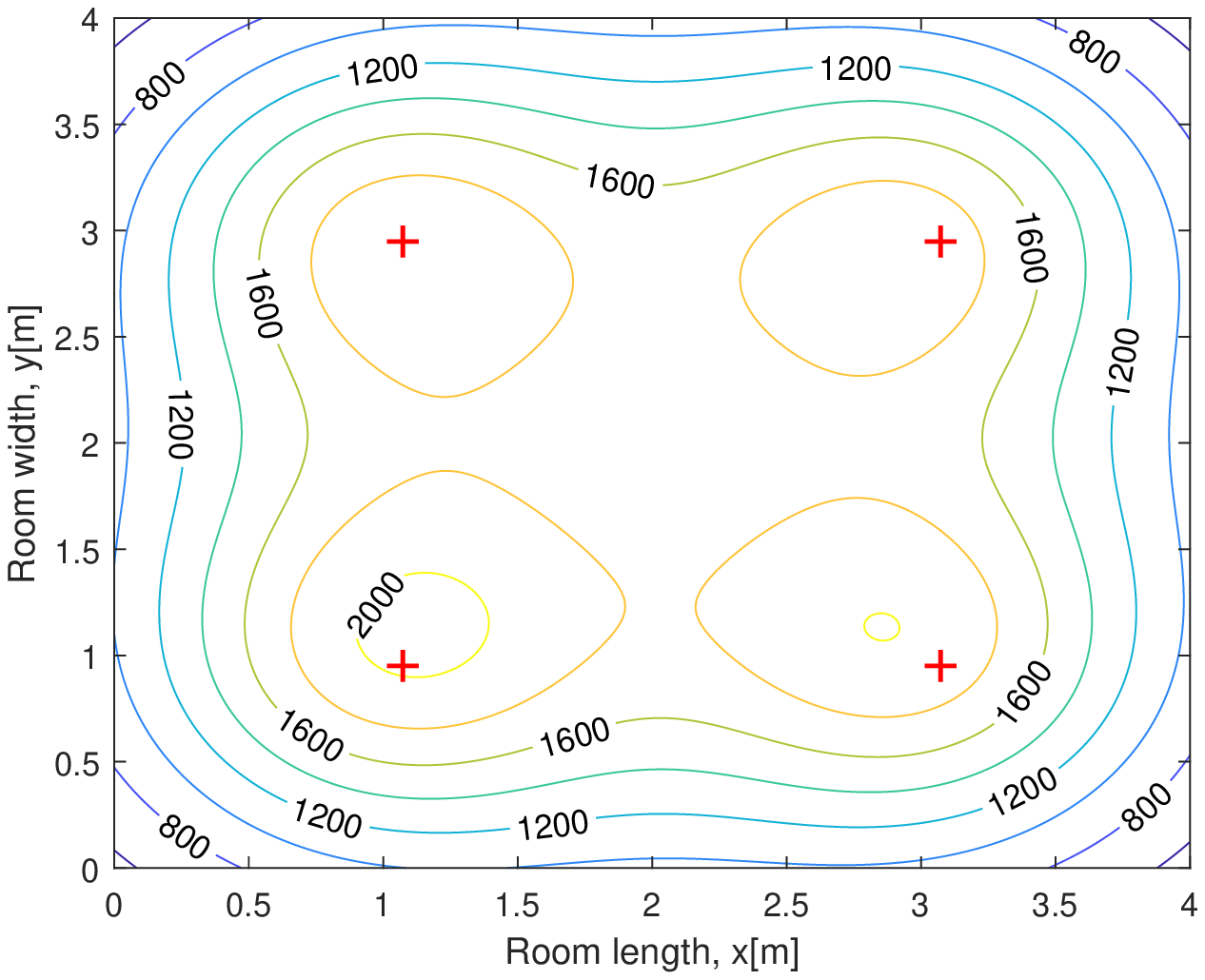}
    }
   \caption{The illuminance distribution on the room at $\eta  = 20\%$. The red crosses in the figure show the position of LEDs. }
   \label{fig:3dcomp}
\end{figure}

To be more clear, the illuminance distribution on the receiver plane at $\eta  = 20\%$ is illustrated in Fig. \ref{fig:3dcomp}. As shown in Fig. \ref{fig:3dcomp}(a), the contour of the illuminance distribution of GDC with traditional sequential selection is uneven.
In particular, two contours around 2400 lx appear under the two LEDs at the bottom of the figure, while that under the two top LEDs are around 1600 lx.
However, Fig. \ref{fig:3dcomp}(b) shows the illuminance distribution of GDC with the proposed incremental algorithm. It is shown that all the four contours under the four LEDs are around 1600 lx. This indicates that the proposed incremental algorithm can achieve a more uniform illumination distribution than the sequential selection.
%\begin{figure}
%\centering
%\includegraphics[width=0.9\textwidth]{comp.eps}
%\caption{Comparison of illumination distribution among optimized GDC, pre-optimized GDC and non-spatial modulation schemes when $\eta  = 20\%$.}
%\label{fig:comp}
%\end{figure}
%

\subsection{BER Evaluations}
\begin{figure}
\centering
\includegraphics[width=0.6\textwidth]{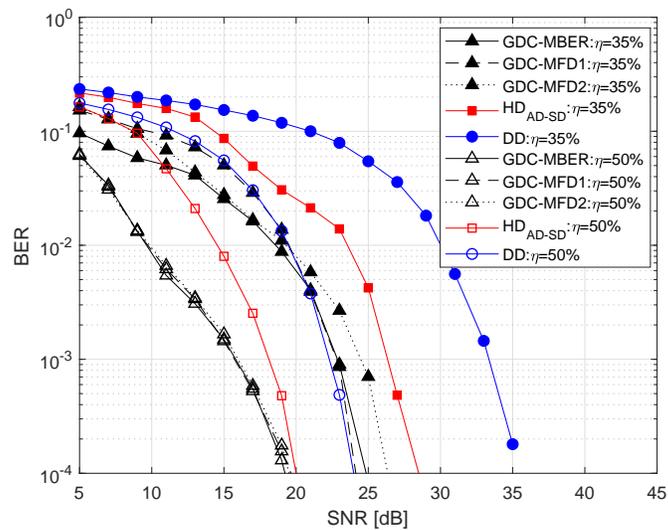}
\caption{Simulated BERs of GDC-MBER, GDC-MFD1, GDC-MFD2 and the baseline schemes (i.e. DD and ${\rm{H}}{{\rm{D}}_{{\rm{AD - SD}}}}$) under 35\% and 50\% dimming levels.}
\label{fig:curve23}
\vspace{-0.5cm}
\end{figure}
Figure \ref{fig:curve23} presents the BERs of GDC-MBER, GDC-MFD1, GDC-MFD2 and the baseline schemes (i.e. HD and DD) at 35\% and 50\% dimming levels.
The simulation results show that GDC-MBER, GDC-MFD1 and GDC-MFD2 have almost the same BER performance at most dimming levels.
This implies that the optimal activation patterns selected by GDC-MFD1 and GDC-MFD2 are the same as that selected by GDC-BER at these dimming levels.
In addition, these three GDC schemes always achieve the best BER performance among all the compared schemes.
For instance, when the dimming level is 50\% and the BER is $10^{-3}$, GDC-MBER, GDC-MFD1 and GDC-MFD2 can yield around 2 dB and 8 dB SNR gains compared to HD and DD, respectively.
We can also observe that the BER performance of GDC is better than that at $\eta = 35\%$.
This is because that more elements are activated in the space-time matrix at $\eta = 50\%$, indicating more information can be transmitted in spatial domain. Therefore, the modulation order of PAM signals can be smaller while maintaining the same data rate, thus resulting in lower BER.

\begin{figure}
\centering
\includegraphics[width=0.6\textwidth]{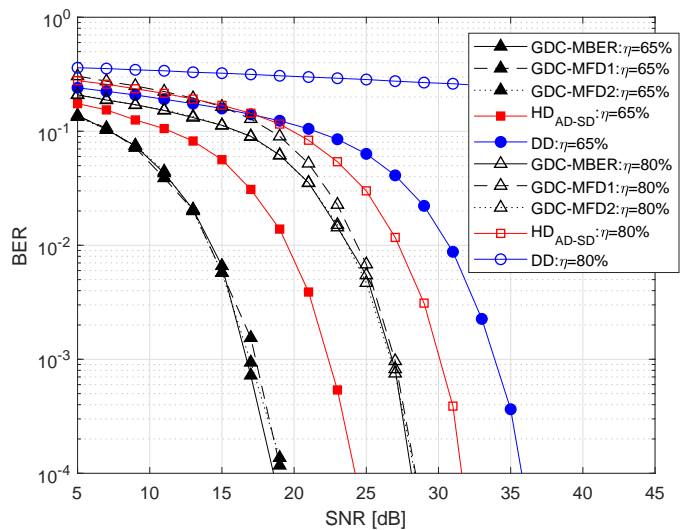}
\caption{Simulated BERs of GDC-MBER, GDC-MFD1, GDC-MFD2 and the baseline schemes under 65\% and 80\% dimming levels.}
\label{fig:curve56}
\vspace{-0.6cm}
\end{figure}
Figure \ref{fig:curve56} illustrates the BERs of all the schemes at 65\% and 80\% dimming levels.
As expected, GDC can still yield BER performance gains compared with HD and DD.
Specifically, when the dimming level is 65\% and BER is ${10^{ - 4}}$, GDC-MBER, GDC-MFD1 and GDC-MFD2 can achieve around 5 dB and 15 dB SNR gains compared with HD and DD schemes, respectively.
It can also be observed that DD suffers significant performance degradation at high dimming levels. This is because that the time period used for transmission is low to achieve the required high dimming level \cite{DDr1}. Therefore, high modulation orders are required to maintain the same data rate. In addition, we can observe that of BERs at 65\% is lower than that at 80\% for the three GDC schemes. This is because when the dimming level is at 80\%, more elements in the space-time matrix must be activated, indicating that less information is transmitted through the acitvation pattern of the space-time matrix of GDC. Therefore, higher modulation order PAM signals must be adopted to maintain the same data rate, resulting in higher BERs. Please also note that in Fig. \ref{fig:curve23} the high dimming level at 50\% has lower BER, while in Fig. \ref{fig:curve56}, the high dimming level at 80\% has higher BER. This is because the bits transmitted through the variation of space-time matrix are ${p_1} = \left\lfloor {{{\log }_2}\left( {{\rm{C}}({N_t}T,{N_S})} \right)} \right\rfloor$, and it increases as $N_S$ becomes moderately large but decreases when $N_S$ becomes excessively large.

%s
\begin{figure}
\centering
\includegraphics[width=0.6\textwidth]{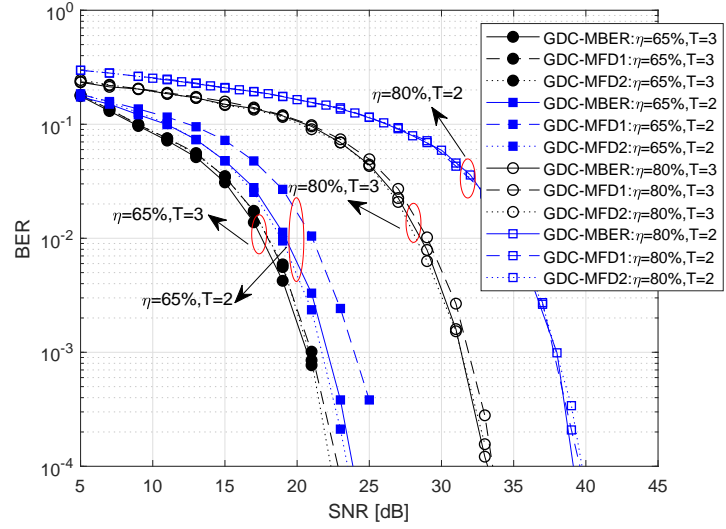}
\caption{Simulation results of BER versus different number of $N_S$ under various dimming levels.}
\label{fig:optimal}
\end{figure}
Figure 6 further simulates the BERs of GDC-MBER, GDC-MFD1 and GDC-MFD2 having different space-time matrix sizes at various dimming levels. In particular, the three GDC schemes with $N_{T} \times T$  space-time matrices of $4 \times 2$  and $4 \times 3$ sizes for $R = 9\;{\rm{bits/symbol}}$ are considered.
The simulation results show that when the dimming level is 80\%, all the three GDC schemes can obtain almost the same performance, indicating the obtained optimal values of $N_S$ are the same in these three schemes.
In contrast, when the dimming level is 65\% and $T=3$, we find that the BERs of GDC-MFD1 and GDC-MFD2 have observable performance gap when compared with that of GDC-MBER.
This is due to the fact that the configuration that achieves the maximum free distance dose not guarantee  minimizing the BER bound in (\ref{aproxBER}), and a minimum BER also depends on the number of error bits for the error events having the free distance.
However, considering the low complexity of GDC-MFD1 and GDC-MFD2, they can be deemed as efficient suboptimal schemes that approach the performance of GDC-MBER in most cases. Besides, it can be observed that the GDC schemes with $4 \times 3$  space-time matrices can achieve better BER performance than the GDC schemes with $4 \times 2$  space-time matrices. This is because the information transmitted by the space-time index increases with the size of the space-time matrix. Given the same data rate, the PAM modulation order used in the GDC schemes with $4 \times 3$  space-time matrices is lower than that in the GDC schemes with $4 \times 2$  space-time matrices, leading to better BER performance.

\begin{figure}
\centering
\includegraphics[width=0.6\textwidth]{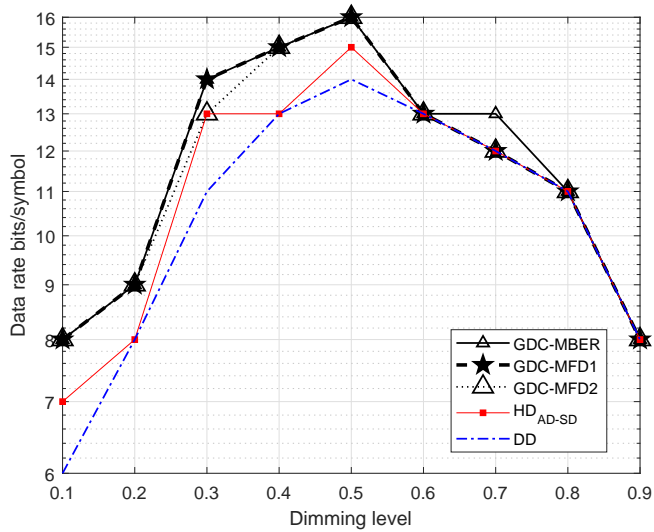}
\caption{Simulation results of transmitted bits versus different dimming levels.}
\label{fig:se}
\end{figure}
Figure \ref{fig:se} evaluates the achievable data rate of the proposed schemes and the compared HD and DD schemes under different illumination levels. The achievable data rate is obtained by adopting the configuration that transmits the maximum number of bits under the BER of $5 \times 10^{-4}$ constraint. As we can observe, GDC-MFD1 and GDC-MFD2 are able to achieve almost the same performance as GDC-MBER. Two special cases can be found. When the dimming level is 0.3, GDC-MFD2 transmits one bit less than GDC-MBER and GDC-MFD1, and when the dimming level is 0.7, GDC-MBER transmits one more bit than GDC-MFD1 and GDC-MFD2. As analyzed in Fig. 6, this is due to the different optimization criterions. We can also observe that the performance gains of the proposed schemes are more observable in low dimming levels such as in 0.1 to 0.4 dimming levels when compared with that in high dimming levels such as in 0.6 to 0.9 dimming levels. This is because as the dimming level constraint increases, more elements in the space-time matrix need to be active to achieve the target dimming level, and this reduces the bits transmitted through the variation of the activation pattern when the dimming level is excessively large. In particular, when all the elements in the space-time matrix have to be active, the proposed GDC schemes degenerate to a AD scheme.

\vspace{-0.cm}
\section{Conclusion}
This work has proposed a generalized dimming control scheme for VLC.  In the proposed dimming control scheme, both the intensity of the transmitted symbols and the number of active elements in the space-time matrix are used to control the dimming level. First, an efficient incremental algorithm for index mapping was proposed to map bits into appropriate activation patterns with uniform illumination. Next, to improve the reliability of GDC with the illumination constraints, we investigated two types of GDC with different criterions.
For GDC-MBER, the analytical BER bound was derived, while the optimal dimming control pattern was obtained by calculating the upper bound of theoretical BER. Furthermore, to reduce the high computational complexity of the exhaustive search of all CPEPs in GDC-MBER, two low-complexity GDC-MFD1 and GDC-MFD2 were proposed.
Specifically, GDC-MFD1 reduced the computational complexity by deriving a lower bound of the free distance based on Rayleigh-Ritz theorem, GDC-MFD2 further utilized the static characteristic of the channel to reduce the computational complexity. The simulation results showed that the proposed GDC-MBER, GDC-MFD1 and GDC-MFD2 can obtain more than 5 dB SNR gains compared with conventional dimming control schemes for a BER of ${10^{ - 4}}$ and a dimming level of 65\% with uniform illumination distribution, which proves that the proposed GDC schemes are promising for future indoor VLC.

\vspace{-0.cm}
\begin{appendices}
\section{$Proof\ of\ Lemma 1$}
\begin{proof}
For a given $N_S$, the value of $M$ can be obtained according to (\ref{abcdef}c). Furthermore, a noise value larger than $\lambda /2$ can cause detection error for a fixed $M$. Hence, minimizing system BER is equivalent to maximizing the value of $\lambda/2$, and  problem (\ref{abcdef}) can be written as
\begin{align}\label{opt-problem1}
&\mathop {\max }\limits_{\lambda ,{B_{\rm{L}}}} \;\;\;{\lambda /2}\\
&\rm{s.\;t.}\;\;\scalebox{1}{${s_m}=\lambda m + {B_{\rm{L}}}$}\tag{\theequation a}\\
&\scalebox{1}{$\;\;\;\;\;\;\;\lambda  + {B_{\rm{L}}} \ge {I_{\rm{L}}}$} \tag{\theequation b}\\
&\scalebox{1}{$\;\;\;\;\;\;\;\lambda M + {B_{\rm{L}}} \le {I_{\rm{H}}}$} \tag{\theequation c}\\
&\scalebox{1}{$\;\;\;\;\;\;\;{\eta _e} = \frac{{\eta {N_t}T}}{{{N_S}}}{\rm{ = }}\frac{{I - {I_{\rm{L}}}}}{{{I_{\rm{H}}} - {I_{\rm{L}}}}}$} \tag{\theequation d}\\
&\scalebox{1}{$\;\;\;\;\;\;\;\sum\limits_{m = {\rm{1}}}^M {{s_m}} = MI$} \tag{\theequation e}.
\end{align}

We can derive $\lambda  \le 2\frac{{I - {I_{\rm{L}}}}}{{M - 1}}$ according to (\ref{opt-problem1}a), (\ref{opt-problem1}c) and (\ref{opt-problem1}e). Similarly, we can derive $\lambda  \le 2\frac{{{I_{\rm{H}}} - I}}{{M - 1}}$ according to (\ref{opt-problem1}b), (\ref{opt-problem1}c) and (\ref{opt-problem1}e). Therefore, we can simplify (\ref{opt-problem1}) as
\begin{eqnarray}\label{opt-problem2}
\begin{split}
&\mathop {\max }\limits_{\lambda ,{B_{\rm{L}}}} \;\;\;\lambda /2\\
&{\rm{s}}{\rm{.t}}{\rm{.}}\;\;\;\lambda  \le \min \left\{ {2\frac{{I - {I_{\rm{L}}}}}{{M - 1}},2\frac{{{I_{\rm{H}}} - I}}{{M - 1}}} \right\}.
\end{split}
\end{eqnarray}
Hence, when $2\frac{{I - {I_{\rm{L}}}}}{{M - 1}} \le 2\frac{{{I_{\rm{H}}} - I}}{{M - 1}}$ (i.e. $I \le \frac{{{I_{\rm{H}}} - {I_{\rm{L}}}}}{2}$), we have $\lambda  = 2\frac{{I - {I_{\rm{L}}}}}{{M - 1}}$. Similarly, when $2\frac{{I - {I_{\rm{L}}}}}{{M - 1}} > 2\frac{{{I_{\rm{H}}} - I}}{{M - 1}}$ (i.e. $I > \frac{{{I_{\rm{H}}} - {I_{\rm{L}}}}}{2}$), we have $\lambda  = 2\frac{{{I_{\rm{H}}} - I}}{{M - 1}}$. Moreover, with the constraints of (\ref{opt-problem1}d) and (\ref{opt-problem1}e), we can calculate the expressions of optimal $\lambda$ and ${B_{\rm{L}}}$ as (\ref{oplam}) and (\ref{opbl}), respectively.
The proof is completed.
\end{proof}
\vspace{-0.2cm}
\end{appendices}
 -------------------------------------------------------------------------
\bibliographystyle{IEEEbib}
\bibliography{stimreference}
\end{document}